\documentclass[10pt, conference]{article}
\usepackage[a4paper, left=2cm, right=2cm, top=2.5cm, bottom=2.5cm]{geometry}

\usepackage{graphicx}
\usepackage{amsmath}
\usepackage{amssymb}
\usepackage{verbatim}
\usepackage{multirow}
\usepackage{tikz}
\usetikzlibrary{quantikz2}
\usepackage{float}
\usepackage{subcaption}
\usepackage{array}
\usepackage{pgfplots}
\pgfplotsset{compat=1.18}
\usepackage{amsthm}
\usepackage{url}
\usepackage{hyperref}
\usepackage{xcolor}
\hypersetup{colorlinks=false,
allbordercolors=white}
\usepackage{algorithmic}
\usepackage{algorithm}
\usepackage{authblk}

\usepackage{cite}
\usepackage{amsmath,amssymb,amsfonts}
\usepackage{algorithmic}
\usepackage{graphicx}  
\usepackage{subcaption}
\usepackage{textcomp}
\usepackage{xcolor}
\usepackage{tikz}
\usepackage{array}
\usepackage{multirow}
\usetikzlibrary{quantikz2}
\usepackage{pgfplots}
\usepackage{caption}
\pgfplotsset{compat=1.18}
\usepgfplotslibrary{groupplots}
\usepackage{pgfplotstable}
\usepackage{placeins}
\def\BibTeX{{\rm B\kern-.05em{\sc i\kern-.025em b}\kern-.08em
    T\kern-.1667em\lower.7ex\hbox{E}\kern-.125emX}}

\title{Impact of Single Rotations and Entanglement Topologies in Quantum Neural Networks}

\author{Marco Mordacci and Michele Amoretti}

\affil{Quantum Software Laboratory, Department of Engineering and Architecture, University of Parma, 43124 Parma, Italy \\
{\texttt{marco.mordacci1@unipr.it}}}

\date{} 

\begin{document}
\maketitle

\begin{abstract}
In this work, an analysis of the performance of different Variational Quantum Circuits is presented, investigating how it changes with respect to entanglement topology, adopted gates, and Quantum Machine Learning tasks to be performed. The objective of the analysis is to identify the optimal way to construct circuits for Quantum Neural Networks. 
In the presented experiments, two types of circuits are used: one with alternating layers of rotations and entanglement, and the other, similar to the first one, but with an additional final layer of rotations. As rotation layers, all combinations of one and two rotation sequences are considered. Four different entanglement topologies are compared: linear, circular, pairwise, and full. Different tasks are considered, namely the generation of probability distributions and images, and image classification. 
Achieved results are correlated with the expressibility and entanglement capability of the different circuits to understand how these features affect performance. 

\end{abstract}

\section{Introduction}

The ability of a variational quantum circuit (VQC) to explore the Hilbert space is measured through the concept of \emph{expressibility}~\cite{sim2019expressibility}. Another important metric is the \emph{entanglement capability}~\cite{sim2019expressibility} of the VQC, which measures the ability of the circuit to generate entangled states among qubits. However, high expressibility and high entanglement can lead to barren plateaus~\cite{holmes2022connecting, ortiz2021entanglement}.

Research about the effect of entanglement has found relevance in VQC~\cite{kim2022entanglement}.
In~\cite{hubregtsen2021evaluation}, the authors found that expressibility has a strong correlation with the results achieved, while entanglement capability shows a weaker correlation.
 In~\cite{ballarin2023entanglement}, the entanglement production of VQCs is analyzed, and the correlation between expressibility and entanglement is discussed. In \cite{correr2024characterizing}, an analysis of entanglement and expressibility of different entanglement topologies was performed.
 The analysis presented in this work goes beyond the scope of~\cite{correr2024characterizing}, providing a broader investigation of entanglement and expressibility across a large set of circuit architectures.
 
This paper provides the following new contributions: 1) a detailed analysis on the performance of three different tasks, the generation of probability distributions and images and the classification of images, by varying the gates and the entanglement topologies used; 2) a thorough analysis of the performance achieved with respect to entanglement and expressibility values in order to guide the design of VQCs; 3) the execution of the best-performing circuits on real quantum hardware to assess their performance.

The remainder of the paper is organized as follows. In Section~\ref{preliminaries}, the tested circuits, the different metrics, and tasks are presented. In Section~\ref{results_sim}, the results obtained in the different tasks and the evaluated metrics are presented and discussed. In Section~\ref{results_ibm}, the performance is assessed on real IBM hardware. Finally, Section~\ref{conclusion} concludes the paper with a summary of the main results and a discussion of future work.

\section{Preliminaries}
\label{preliminaries}

\begin{figure}[htpb]
    \centering
    \begin{minipage}{0.14\textwidth}
        \centering
        \begin{subfigure}[b]{\textwidth}
            \centering
            \includegraphics[width=\textwidth]{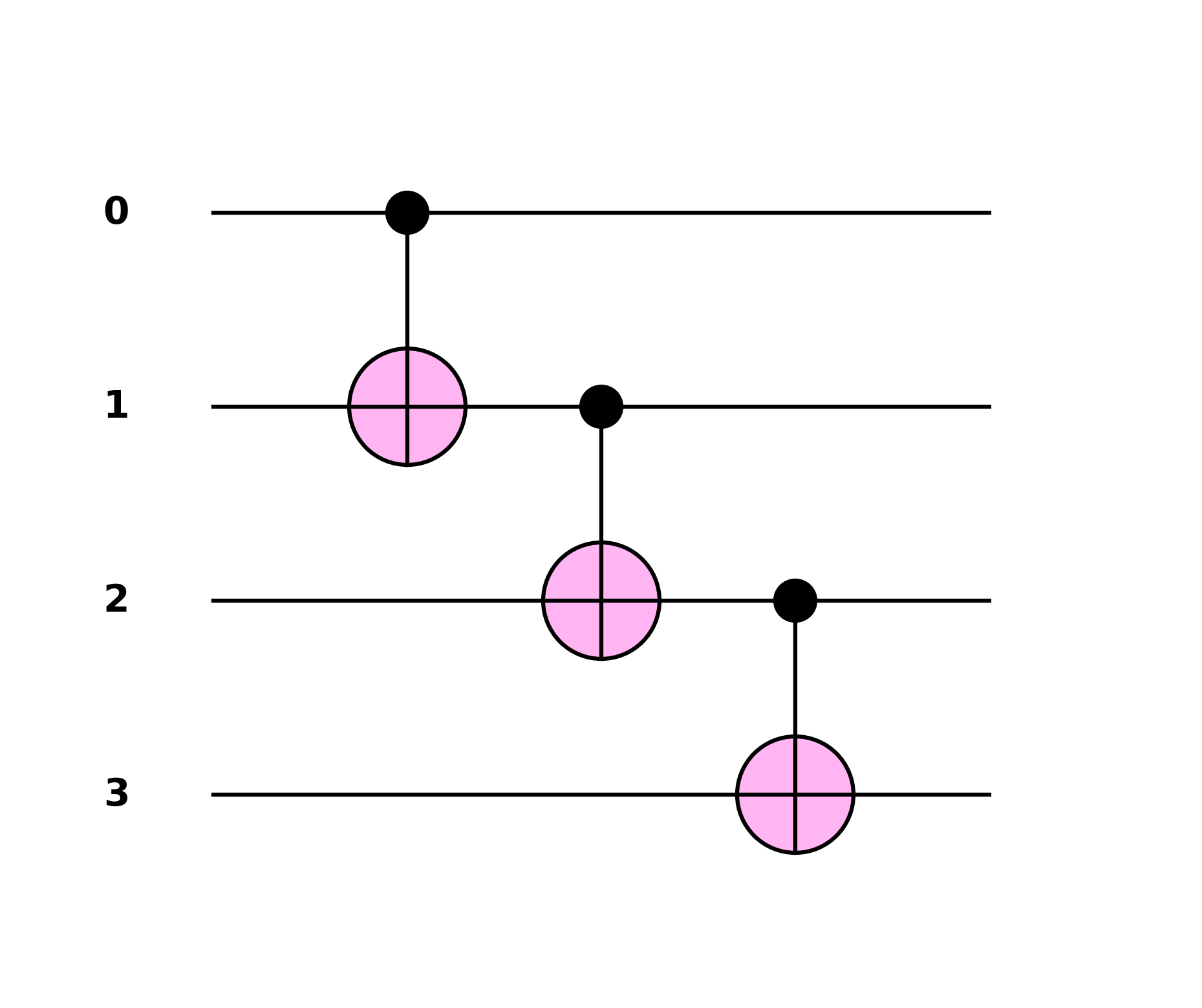}
            \caption{Linear}
            \label{fig:lin}
        \end{subfigure}
        \vfill
        \begin{subfigure}[b]{\textwidth}
            \centering
            \includegraphics[width=\textwidth]{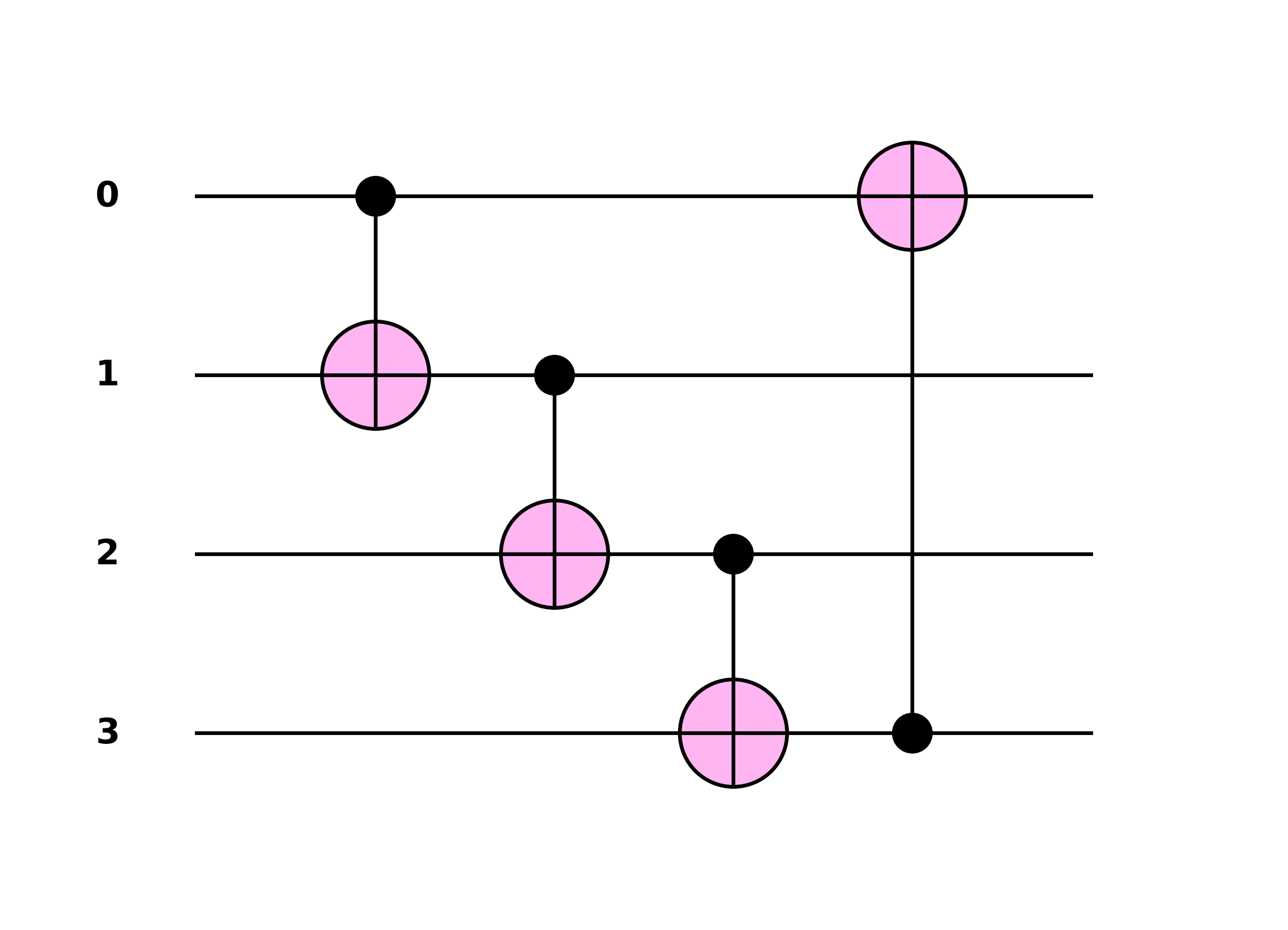}
            \caption{Circular}
            \label{fig:circular}
        \end{subfigure}
    \end{minipage}
    \begin{minipage}{0.14\textwidth}
        \centering
        \begin{subfigure}[b]{\textwidth}
            \centering
            \includegraphics[width=\textwidth]{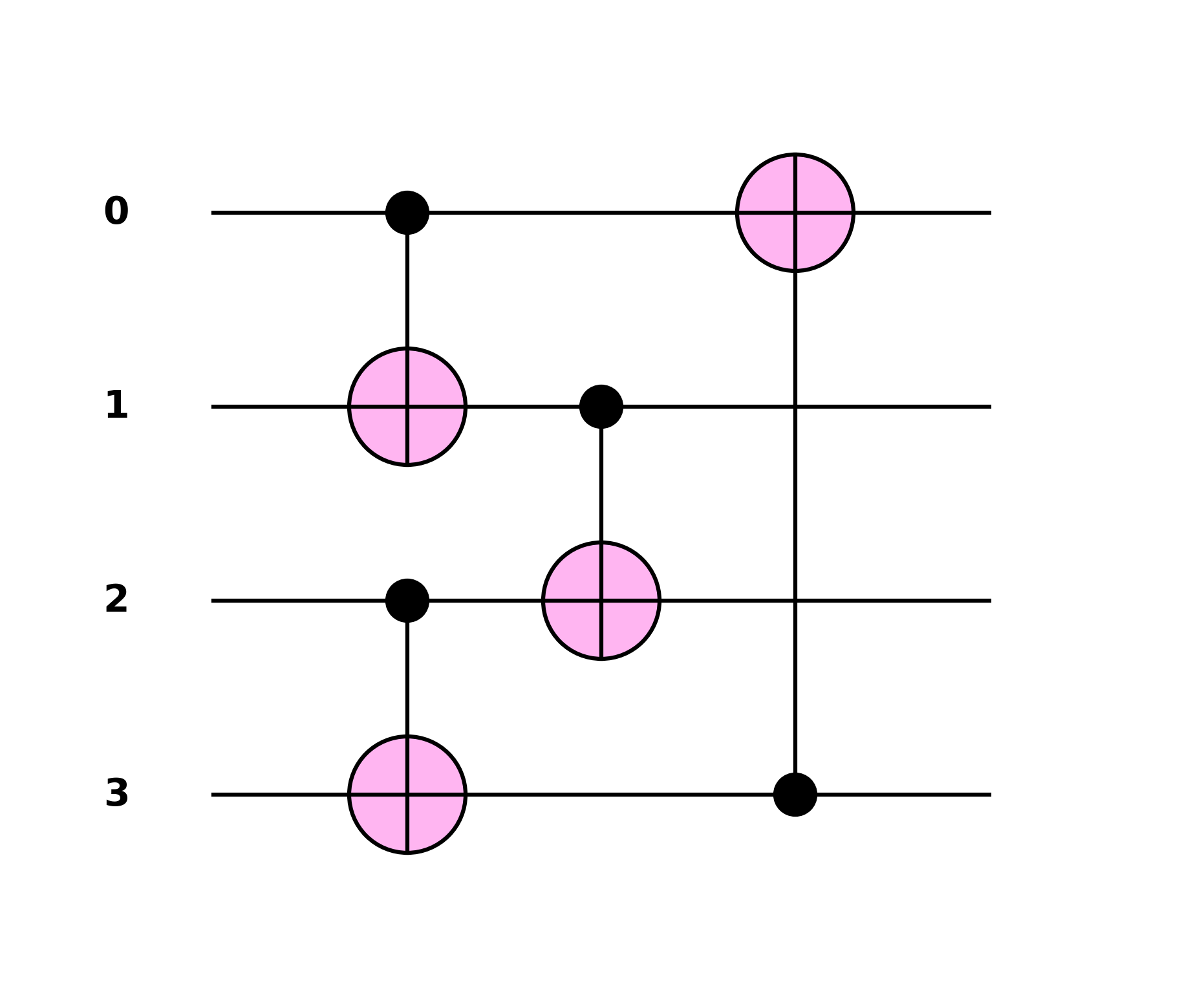}
            \caption{Pairwise}
            \label{fig:pairwise}
        \end{subfigure}
        \vfill
        \begin{subfigure}[b]{\textwidth}
            \centering
            \includegraphics[width=\textwidth]{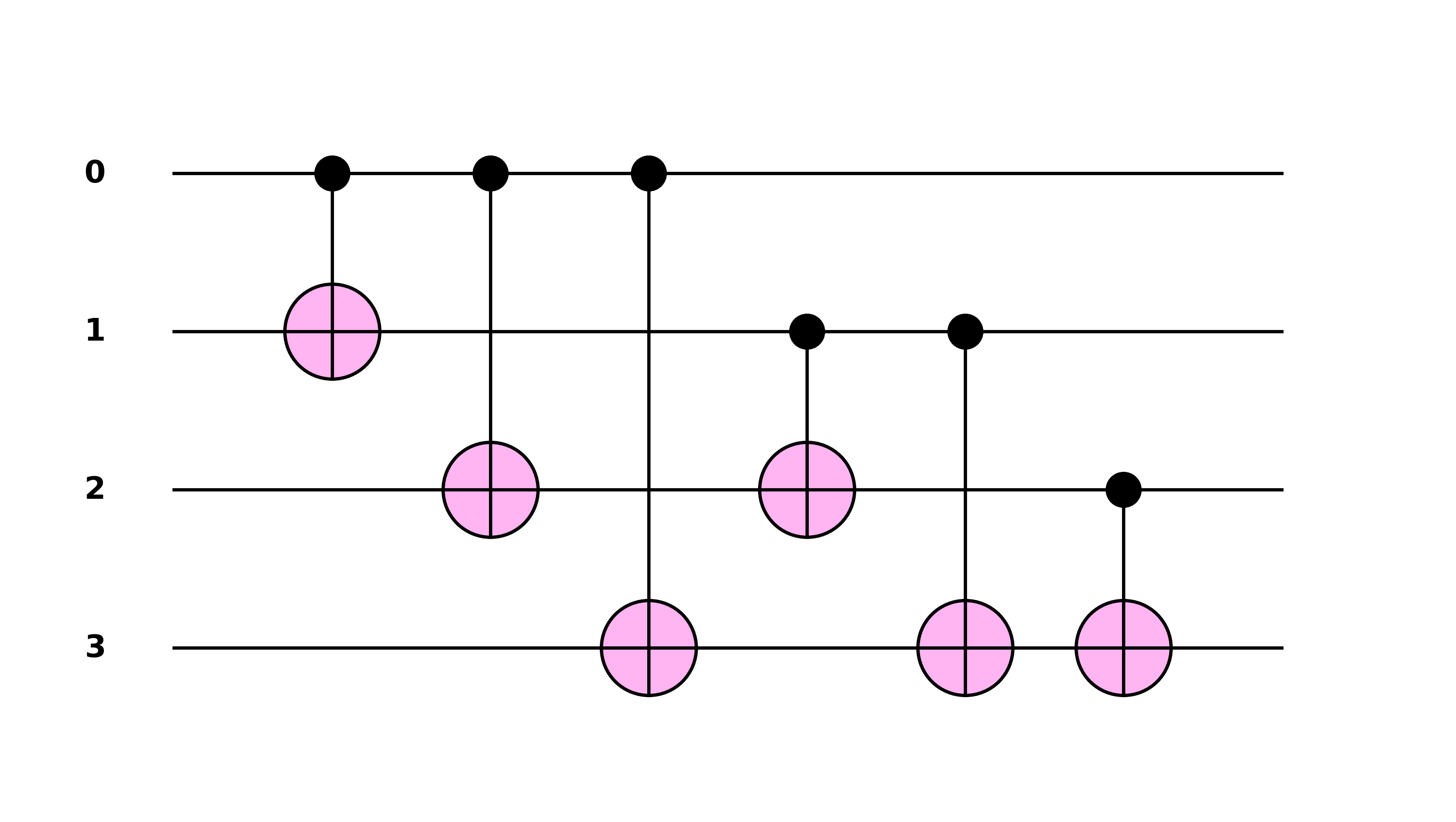}
            \caption{Full}
            \label{fig:full}
        \end{subfigure}
    \end{minipage}

    \caption{The considered entanglement topologies.}
    \label{fig:entanglement}
\end{figure}

Two circuit types are considered to investigate how performance is affected by changes in the entanglement topology, the number of circuit layers, the different gates used, and depending on the task to be performed.
The first circuit, named $C_1$, consists of alternating layers of rotations and entanglement, while the second circuit, named $C_2$, follows the same structure as $C_1$ but with an additional layer of rotations at the end. 
Four different entanglement topologies, shown in Fig.~\ref{fig:entanglement}, are considered: linear, circular, full, and pairwise (CNOT is used as entanglement generator). The considered circuits use 6 qubits because this number of qubits is used for the different tasks tested.

Three tasks are evaluated: the generation of random probability distributions, the generation of images, and the classification of images.
The generative tasks are tested with a Quantum Generative Adversarial Network (QGAN) composed of a quantum generator and a classical discriminator, following the structure proposed in~\cite{zoufal2019quantum} for generation of probability distributions, and~\cite{huang2021experimental} for the generation of images. Each circuit is tested with 10 random probability distributions, with performance measured by the Hellinger distance between the original and generated distributions. 

Image generation is tested using the MNIST dataset, and evaluated by the Fr\'echet Inception Distance (FID). Additionally, the MNIST dataset is employed for image classification, considering 2 (classes 0-1), 4 (classes 0-3), and 6-class (classes 0-5) classification tasks. Accuracy is used as the evaluation metric.
Optimal performance corresponds to lower FID and Hellinger distance, and higher accuracy.

The expressibility, introduced in~\cite{sim2019expressibility}, is calculated as:
\begin{equation}
    expr = D_{KL}(P_{VQC}(F; \theta) || P_{Haar}(F)),
\end{equation}
where $D_{KL}$ is the Kullback-Leibler divergence, $F$ corresponds to the fidelity, $P_{VQC}$ is the estimated probability distribution of fidelities resulting from sampling states from a VQC, and $P_{Haar}(F)$ is the probability density function of fidelities for the ensemble of Haar random states. This measure computes the circuit's ability to generate states that are well representative of the Hilbert space, and it is calculated by comparing the distribution of states obtained from sampling the parameters of a VQC to the uniform distribution, i.e., the ensemble of Haar-random states.
A lower expressibility value indicates a higher level of circuit expressiveness.

The entanglement capability~\cite{sim2019expressibility} is computed as the Meyer-Wallach entanglement measure~\cite{meyer2001global,brennen2003observable}, i.e.,
\begin{equation}
    Q(\ket{\psi}) =  2(1 - \frac{1}{n} \sum_{k=0}^{n-1}Tr[\rho_k^2]),
\end{equation}
where $n$ is the number of qubits. 
The measure satisfies the following properties: 1) $0 \leq Q(\ket{\psi} \leq 1$; 2)$Q(\ket{\psi}) = 0 $ iff $\ket{\psi}$ is a product state; 3) $Q(\ket{\psi})$ is invariant under local unitaries.

\section{Simulation Results and Analysis of Entanglement and Expressibility}
\label{results_sim}
Simulation tests are performed using a Linux machine equipped with an AMD EPYC 7282 CPU and 256 GB RAM.

In all simulations, the combinations of the 3 rotations are not considered since using all of them does not improve the performance. The same results can be produced just by the combination of 2 rotations, as shown in Table~\ref{tab:comparison2_3rots}. This phenomenon likely arises from the fact that the third rotation's effect can be obtained through a combination of the first two rotations.

\begin{table}[htbp]
\centering
\resizebox{0.5\textwidth}{!}{
\renewcommand{\arraystretch}{1.2}
    \begin{tabular}{|>{\centering\arraybackslash}m{1cm}|>{\centering\arraybackslash}m{0.8cm}|>{\centering\arraybackslash}m{0.8cm}|>{\centering\arraybackslash}m{1cm}|>{\centering\arraybackslash}m{1cm}|>{\centering\arraybackslash}m{1cm}|>{\centering\arraybackslash}m{1cm}|>{\centering\arraybackslash}m{1cm}|}
     \hline
     Circuit & $\#$ layer & $\#$ rotations & Hellinger & FID & Acc 2C & Acc 4C & Acc 6C \\
     \hline
     \multirow{4}{*}{\centering $C_1$} &\multirow{2}{*}{\centering 1} & 2&$0.31$&$100$&$99\%$&$61\%$&$41\%$ \\
     \cline{3-8}
     & &3 &$0.31$ &$105$&$99\%$&$61\%$&$41\%$\\
     \cline{2-8}
     &\multirow{2}{*}{\centering 2} &2 &$0.27$&$45$&$99\%$&$78\%$&$59\%$ \\
     \cline{3-8}
     & &3 &$0.27$&$45$&$99\%$&$77\%$&$60\%$ \\
     \cline{1-8}
     \multirow{4}{*}{\centering $C_2$} & \multirow{2}{*}{\centering 1} &2 &$0.24$&$55$&$99\%$&$76\%$&$55\%$ \\
     \cline{3-8}
      & &3 &$0.25$ &$55$&$99\%$&$76\%$&$55\%$\\
     \cline{2-8}
     &\multirow{2}{*}{\centering 2} &2 &$0.28$&$40$&$99\%$&$84\%$&$68\%$ \\
     \cline{3-8}
     & &3 &$0.28$&$40$&$99\%$&$85\%$&$70\%$ \\
     \hline
     
    \end{tabular}
    }
    \caption{Comparison between the best performance of the combinations of two and three rotations.}
    \label{tab:comparison2_3rots}
\end{table}

In~\cite{correr2024characterizing}, the authors take into account only the results of $R_x$ followed by $R_y$. In this work, all possible combinations of $1$ and $2$ rotations are considered to understand how the different rotations affect the expressibility and entanglement values. 

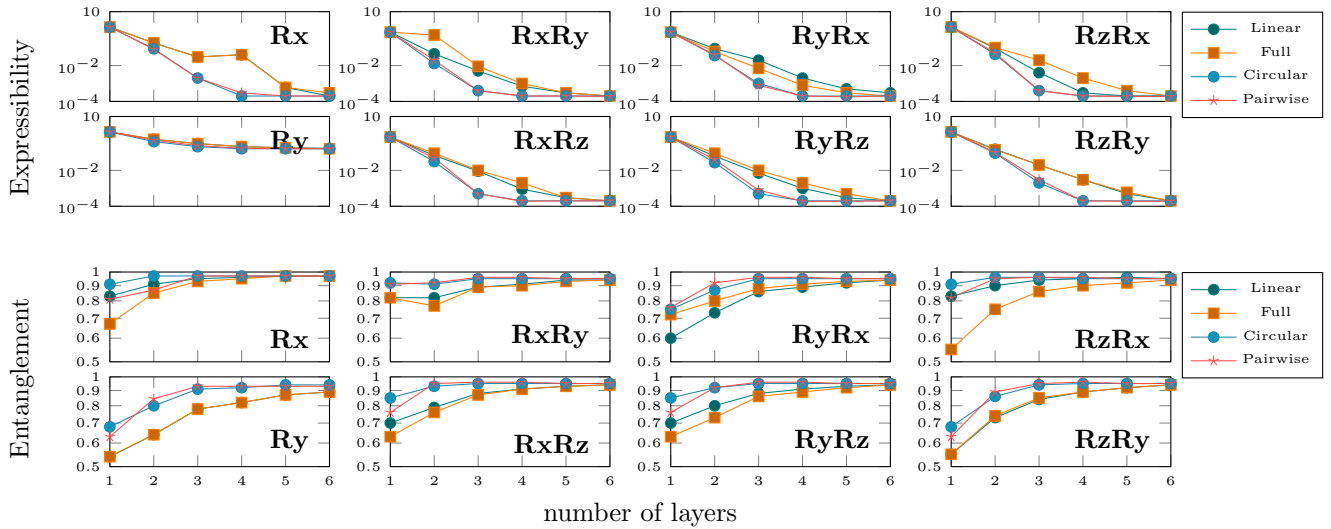
\begin{figure*}[htbp]
    \centering
    \begin{subfigure}{\textwidth}
        \centering
        \begin{tikzpicture}
            \begin{groupplot}[
                group style={
                    group size=4 by 2,
                    vertical sep=0.2cm,
                    horizontal sep=0.8cm,  
                    group name=myplots,
                },
                width=0.17\columnwidth,    
                height=0.07\columnwidth,
                scale only axis,
                title={},
                xmin=1, xmax=6,
                xtick={1, 2, 3, 4, 5, 6},
                samples at={1, 2, 3, 4, 5, 6},
                ymode=log,
                ymin=1e-4, ymax=1e1,
                ytick={1e1,1e-2,1e-4},
                yticklabels={$10$, $10^{-2}$, $10^{-4}$},
                cycle list name=exotic, 
                legend style={font=\tiny, at={(1.05,1)}, anchor=north west},
                ticklabel style={font=\tiny},
            ]

            \nextgroupplot[xticklabels=\empty]
                \addplot table[x=layer, y=lin] {\rxexp};
                \addplot table[x=layer, y=full] {\rxexp};
                \addplot table[x=layer, y=circ] {\rxexp};
                \addplot table[x=layer, y=pw] {\rxexp};
                \node[anchor=north east, font=\bfseries] at (rel axis cs:0.95, 0.95) {Rx};

            \nextgroupplot[xticklabels=\empty]
                \addplot table[x=layer, y=lin] {\rxryexp};
                \addplot table[x=layer, y=full] {\rxryexp};
                \addplot table[x=layer, y=circ] {\rxryexp};
                \addplot table[x=layer, y=pw] {\rxryexp};
                \node[anchor=north east, font=\bfseries] at (rel axis cs:0.95, 0.95) {RxRy};

            \nextgroupplot[xticklabels=\empty]
                \addplot table[x=layer, y=lin] {\ryrxexp};
                \addplot table[x=layer, y=full] {\ryrxexp};
                \addplot table[x=layer, y=circ] {\ryrxexp};
                \addplot table[x=layer, y=pw] {\ryrxexp};
                \node[anchor=north east, font=\bfseries] at (rel axis cs:0.95, 0.95) {RyRx};
            
            \nextgroupplot[xticklabels=\empty] 
                \addplot table[x=layer, y=lin] {\rzrxexp}; \addlegendentry{Linear}
                \addplot table[x=layer, y=full] {\rzrxexp}; \addlegendentry{Full}
                \addplot table[x=layer, y=circ] {\rzrxexp}; \addlegendentry{Circular}
                \addplot table[x=layer, y=pw] {\rzrxexp}; \addlegendentry{Pairwise}
                \node[anchor=north east, font=\bfseries] at (rel axis cs:0.95, 0.95) {RzRx};
            
             \nextgroupplot[xticklabels=\empty]
                \addplot table[x=layer, y=lin] {\ryexp};
                \addplot table[x=layer, y=full] {\ryexp};
                \addplot table[x=layer, y=circ] {\ryexp};
                \addplot table[x=layer, y=pw] {\ryexp};
                \node[anchor=north east, font=\bfseries] at (rel axis cs:0.95, 0.95) {Ry};

            \nextgroupplot[xticklabels=\empty]
                \addplot table[x=layer, y=lin] {\rxrzexp};
                \addplot table[x=layer, y=full] {\rxrzexp};
                \addplot table[x=layer, y=circ] {\rxrzexp};
                \addplot table[x=layer, y=pw] {\rxrzexp};
                \node[anchor=north east, font=\bfseries] at (rel axis cs:0.95, 0.95) {RxRz};

            \nextgroupplot[xticklabels=\empty]
                \addplot table[x=layer, y=lin] {\ryrzexp};
                \addplot table[x=layer, y=full] {\ryrzexp};
                \addplot table[x=layer, y=circ] {\ryrzexp};
                \addplot table[x=layer, y=pw] {\ryrzexp};
                \node[anchor=north east, font=\bfseries] at (rel axis cs:0.95, 0.95) {RyRz};
            
            \nextgroupplot[xticklabels=\empty]
                \addplot table[x=layer, y=lin] {\rzryexp};
                \addplot table[x=layer, y=full] {\rzryexp};
                \addplot table[x=layer, y=circ] {\rzryexp};
                \addplot table[x=layer, y=pw] {\rzryexp};
                \node[anchor=north east, font=\bfseries] at (rel axis cs:0.95, 0.95) {RzRy};
            \end{groupplot}

            \node at (myplots c1r1.north west) [rotate=90, anchor=south, yshift=2.5em, xshift=-4em] {Expressibility};
            
        \end{tikzpicture}
        \label{fig:expr_c1_sub}
    \end{subfigure}


    \begin{subfigure}{\textwidth}
        \centering
        \begin{tikzpicture}
            \begin{groupplot}[
                group style={
                    group size=4 by 2,
                    vertical sep=0.2cm,
                    horizontal sep=0.8cm,  
                    group name=myplots2,
                },
                width=0.17\columnwidth,    
                height=0.07\columnwidth,
                scale only axis,
                title={},
                xmin=1, xmax=6,
                xtick={1, 2, 3, 4, 5, 6},
                samples at={1, 2, 3, 4, 5, 6},
                ymode=log,
                ymin=0.5, ymax=1,
                ytick={0.5, 0.6, 0.7, 0.8, 0.9, 1},
                yticklabels={$0.5$, $0.6$, $0.7$, $0.8$, $0.9$, $1$},
                cycle list name=exotic,
                legend style={font=\tiny, at={(1.05,1)}, anchor=north west},
                ticklabel style={font=\tiny},
            ]

            \nextgroupplot[xticklabels=\empty]
                \addplot table[x=layer, y=lin] {\rxent};
                \addplot table[x=layer, y=full] {\rxent};
                \addplot table[x=layer, y=circ] {\rxent};
                \addplot table[x=layer, y=pw] {\rxent};
                \node[anchor=south east, font=\bfseries] at (rel axis cs:0.95, 0.05) {Rx};

            \nextgroupplot[xticklabels=\empty]
                \addplot table[x=layer, y=lin] {\rxryent};
                \addplot table[x=layer, y=full] {\rxryent};
                \addplot table[x=layer, y=circ] {\rxryent};
                \addplot table[x=layer, y=pw] {\rxryent};
                \node[anchor=south east, font=\bfseries] at (rel axis cs:0.95, 0.05) {RxRy};

            \nextgroupplot[xticklabels=\empty]
                \addplot table[x=layer, y=lin] {\ryrxent};
                \addplot table[x=layer, y=full] {\ryrxent};
                \addplot table[x=layer, y=circ] {\ryrxent};
                \addplot table[x=layer, y=pw] {\ryrxent};
                \node[anchor=south east, font=\bfseries] at (rel axis cs:0.95, 0.05) {RyRx};
            
            \nextgroupplot[xticklabels=\empty] 
                \addplot table[x=layer, y=lin] {\rzrxent}; \addlegendentry{Linear}
                \addplot table[x=layer, y=full] {\rzrxent}; \addlegendentry{Full}
                \addplot table[x=layer, y=circ] {\rzrxent}; \addlegendentry{Circular}
                \addplot table[x=layer, y=pw] {\rzrxent}; \addlegendentry{Pairwise}
                \node[anchor=south east, font=\bfseries] at (rel axis cs:0.95, 0.05) {RzRx};
            
            \nextgroupplot
                \addplot table[x=layer, y=lin] {\ryent};
                \addplot table[x=layer, y=full] {\ryent};
                \addplot table[x=layer, y=circ] {\ryent};
                \addplot table[x=layer, y=pw] {\ryent};
                \node[anchor=south east, font=\bfseries] at (rel axis cs:0.95, 0.05) {Ry};

            \nextgroupplot
                \addplot table[x=layer, y=lin] {\rxrzent};
                \addplot table[x=layer, y=full] {\rxrzent};
                \addplot table[x=layer, y=circ] {\rxrzent};
                \addplot table[x=layer, y=pw] {\rxrzent};
                \node[anchor=south east, font=\bfseries] at (rel axis cs:0.95, 0.05) {RxRz};

            \nextgroupplot
                \addplot table[x=layer, y=lin] {\ryrzent};
                \addplot table[x=layer, y=full] {\ryrzent};
                \addplot table[x=layer, y=circ] {\ryrzent};
                \addplot table[x=layer, y=pw] {\ryrzent};
                \node[anchor=south east, font=\bfseries] at (rel axis cs:0.95, 0.05) {RyRz};
            
            \nextgroupplot
                \addplot table[x=layer, y=lin] {\rzryent};
                \addplot table[x=layer, y=full] {\rzryent};
                \addplot table[x=layer, y=circ] {\rzryent};
                \addplot table[x=layer, y=pw] {\rzryent};
                \node[anchor=south east, font=\bfseries] at (rel axis cs:0.95, 0.05) {RzRy};
            \end{groupplot}

            \node[anchor=north] at ($(myplots2 c2r2.south)!0.5!(myplots2 c3r2.south)$) [yshift=-1em] {number of layers};
            \node at (myplots2 c1r1.north west) [rotate=90, anchor=south, yshift=2.5em, xshift=-4em] {Entanglement};

        \end{tikzpicture}
        \label{fig:ent_c1_sub}
    \end{subfigure}
    \captionsetup{skip=-0.05pt}
    \caption{Expressibility and Entanglement of different rotation combinations with circuit $C_1$. Expressibility quantified by Kullback-Leibler divergence; lower values mean better expressibility.}
    \label{fig:combined_c1}
\end{figure*}

\begin{figure*}[htb]
    \centering
    \begin{subfigure}{\textwidth}
        \centering
        \begin{tikzpicture}
            \begin{groupplot}[
                group style={
                    group size=4 by 2,
                    vertical sep=0.2cm,
                    horizontal sep=0.8cm,  
                    group name=myplots,
                },
                width=0.17\columnwidth,    
                height=0.07\columnwidth,
                scale only axis,
                title={},
                xmin=1, xmax=6,
                xtick={1, 2, 3, 4, 5, 6},
                samples at={1, 2, 3, 4, 5, 6},
                ymode=log,
                ymin=1e-4, ymax=1e1,
                ytick={1e1,1e-2,1e-4},
                yticklabels={$10$, $10^{-2}$, $10^{-4}$},
                cycle list name=exotic, 
                legend style={font=\tiny, at={(1.05,1)}, anchor=north west},
                ticklabel style={font=\tiny},
            ]

            \nextgroupplot[xticklabels=\empty] 
            \addplot table[x=layer, y=lin] {\rxexpc};
            \addplot table[x=layer, y=full] {\rxexpc};
            \addplot table[x=layer, y=circ] {\rxexpc};
            \addplot table[x=layer, y=pw] {\rxexpc};
            \node[anchor=north east, font=\bfseries] at (rel axis cs:0.95, 0.95) {Rx};

        \nextgroupplot[xticklabels=\empty] 
            \addplot table[x=layer, y=lin] {\rxryexpc};
            \addplot table[x=layer, y=full] {\rxryexpc};
            \addplot table[x=layer, y=circ] {\rxryexpc};
            \addplot table[x=layer, y=pw] {\rxryexpc};
            \node[anchor=north east, font=\bfseries] at (rel axis cs:0.95, 0.95) {RxRy};

        \nextgroupplot[xticklabels=\empty] 
            \addplot table[x=layer, y=lin] {\ryrxexpc};
            \addplot table[x=layer, y=full] {\ryrxexpc};
            \addplot table[x=layer, y=circ] {\ryrxexpc};
            \addplot table[x=layer, y=pw] {\ryrxexpc};
            \node[anchor=north east, font=\bfseries] at (rel axis cs:0.95, 0.95) {RyRx};
            
        \nextgroupplot[xticklabels=\empty] 
            \addplot table[x=layer, y=lin] {\rzrxexpc}; \addlegendentry{Linear}
            \addplot table[x=layer, y=full] {\rzrxexpc}; \addlegendentry{Full}
            \addplot table[x=layer, y=circ] {\rzrxexpc}; \addlegendentry{Circular}
            \addplot table[x=layer, y=pw] {\rzrxexpc}; \addlegendentry{Pairwise}
            \node[anchor=north east, font=\bfseries] at (rel axis cs:0.95, 0.95) {RzRx};
        

        \nextgroupplot[xticklabels=\empty] 
            \addplot table[x=layer, y=lin] {\ryexpc};
            \addplot table[x=layer, y=full] {\ryexpc};
            \addplot table[x=layer, y=circ] {\ryexpc};
            \addplot table[x=layer, y=pw] {\ryexpc};
            \node[anchor=north east, font=\bfseries] at (rel axis cs:0.95, 0.95) {Ry};

        \nextgroupplot[xticklabels=\empty] 
            \addplot table[x=layer, y=lin] {\rxrzexpc};
            \addplot table[x=layer, y=full] {\rxrzexpc};
            \addplot table[x=layer, y=circ] {\rxrzexpc};
            \addplot table[x=layer, y=pw] {\rxrzexpc};
            \node[anchor=north east, font=\bfseries] at (rel axis cs:0.95, 0.95) {RxRz};

        \nextgroupplot[xticklabels=\empty] 
            \addplot table[x=layer, y=lin] {\ryrzexpc};
            \addplot table[x=layer, y=full] {\ryrzexpc};
            \addplot table[x=layer, y=circ] {\ryrzexpc};
            \addplot table[x=layer, y=pw] {\ryrzexpc};
            \node[anchor=north east, font=\bfseries] at (rel axis cs:0.95, 0.95) {RyRz};
            
        \nextgroupplot[xticklabels=\empty] 
            \addplot table[x=layer, y=lin] {\rzryexpc};
            \addplot table[x=layer, y=full] {\rzryexpc};
            \addplot table[x=layer, y=circ] {\rzryexpc};
            \addplot table[x=layer, y=pw] {\rzryexpc};
            \node[anchor=north east, font=\bfseries] at (rel axis cs:0.95, 0.95) {RzRy};
            
        \end{groupplot}

            \node at (myplots c1r1.north west) [rotate=90, anchor=south, yshift=2.5em, xshift=-4em] {Expressibility};
            
        \end{tikzpicture}
        \label{fig:expr_c2_sub}
    \end{subfigure}


    \begin{subfigure}{\textwidth}
        \centering
        \begin{tikzpicture}
            \begin{groupplot}[
                group style={
                    group size=4 by 2,
                    vertical sep=0.2cm,
                    horizontal sep=0.8cm,  
                    group name=myplots2,
                },
                width=0.17\columnwidth,    
                height=0.07\columnwidth,
                scale only axis,
                title={},
                xmin=1, xmax=6,
                xtick={1, 2, 3, 4, 5, 6},
                samples at={1, 2, 3, 4, 5, 6},
                ymode=log,
                ymin=0.5, ymax=1,
                ytick={0.5, 0.6, 0.7, 0.8, 0.9, 1},
                yticklabels={$0.5$, $0.6$, $0.7$, $0.8$, $0.9$, $1$},
                cycle list name=exotic,
                legend style={font=\tiny, at={(1.05,1)}, anchor=north west},
                ticklabel style={font=\tiny},
            ]

        
        \nextgroupplot[xticklabels=\empty] 
            \addplot table[x=layer, y=lin] {\rxentc};
            \addplot table[x=layer, y=full] {\rxentc};
            \addplot table[x=layer, y=circ] {\rxentc};
            \addplot table[x=layer, y=pw] {\rxentc};
            \node[anchor=south east, font=\bfseries] at (rel axis cs:0.95, 0.05) {Rx};

        \nextgroupplot[xticklabels=\empty] 
            \addplot table[x=layer, y=lin] {\rxryentc};
            \addplot table[x=layer, y=full] {\rxryentc};
            \addplot table[x=layer, y=circ] {\rxryentc};
            \addplot table[x=layer, y=pw] {\rxryentc};
            \node[anchor=south east, font=\bfseries] at (rel axis cs:0.95, 0.05) {RxRy};

        \nextgroupplot[xticklabels=\empty] 
            \addplot table[x=layer, y=lin] {\ryrxentc};
            \addplot table[x=layer, y=full] {\ryrxentc};
            \addplot table[x=layer, y=circ] {\ryrxentc};
            \addplot table[x=layer, y=pw] {\ryrxentc};
            \node[anchor=south east, font=\bfseries] at (rel axis cs:0.95, 0.05) {RyRx};
            
        \nextgroupplot[xticklabels=\empty] 
            \addplot table[x=layer, y=lin] {\rzrxentc}; \addlegendentry{Linear}
            \addplot table[x=layer, y=full] {\rzrxentc}; \addlegendentry{Full}
            \addplot table[x=layer, y=circ] {\rzrxentc}; \addlegendentry{Circular}
            \addplot table[x=layer, y=pw] {\rzrxentc}; \addlegendentry{Pairwise}
            \node[anchor=south east, font=\bfseries] at (rel axis cs:0.95, 0.05) {RzRx};
        

        \nextgroupplot 
            \addplot table[x=layer, y=lin] {\ryentc};
            \addplot table[x=layer, y=full] {\ryentc};
            \addplot table[x=layer, y=circ] {\ryentc};
            \addplot table[x=layer, y=pw] {\ryentc};
            \node[anchor=south east, font=\bfseries] at (rel axis cs:0.95, 0.05) {Ry};

        \nextgroupplot 
            \addplot table[x=layer, y=lin] {\rxrzentc};
            \addplot table[x=layer, y=full] {\rxrzentc};
            \addplot table[x=layer, y=circ] {\rxrzentc};
            \addplot table[x=layer, y=pw] {\rxrzentc};
            \node[anchor=south east, font=\bfseries] at (rel axis cs:0.95, 0.05) {RxRz};

        \nextgroupplot 
            \addplot table[x=layer, y=lin] {\ryrzentc};
            \addplot table[x=layer, y=full] {\ryrzentc};
            \addplot table[x=layer, y=circ] {\ryrzentc};
            \addplot table[x=layer, y=pw] {\ryrzentc};
            \node[anchor=south east, font=\bfseries] at (rel axis cs:0.95, 0.05) {RyRz};
            
        \nextgroupplot 
            \addplot table[x=layer, y=lin] {\rzryentc};
            \addplot table[x=layer, y=full] {\rzryentc};
            \addplot table[x=layer, y=circ] {\rzryentc};
            \addplot table[x=layer, y=pw] {\rzryentc};
            \node[anchor=south east, font=\bfseries] at (rel axis cs:0.95, 0.05) {RzRy};
            
        \end{groupplot}

            \node[anchor=north] at ($(myplots2 c2r2.south)!0.5!(myplots2 c3r2.south)$) [yshift=-1em] {number of layers};
            \node at (myplots2 c1r1.north west) [rotate=90, anchor=south, yshift=2.5em, xshift=-4em] {Entanglement};

        \end{tikzpicture}
        \label{fig:ent_c2_sub}
    \end{subfigure}
    \captionsetup{skip=-0.05pt}
    \caption{Expressibility and Entanglement of different rotation combinations with circuit $C_2$. Expressibility quantified by Kullback-Leibler divergence; lower values mean better expressibility.}
    \label{fig:combined_c2}
\end{figure*}
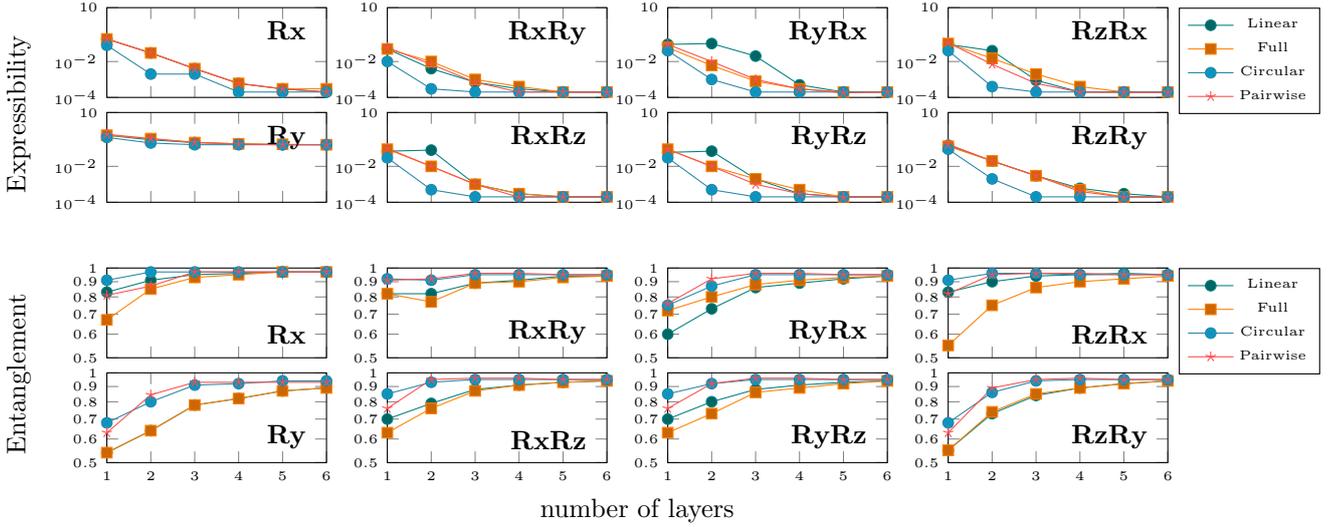

In Figures~\ref{fig:combined_c1} and~\ref{fig:combined_c2}, the values of the entanglement capability and expressibility are shown. The results illustrate a similar trend for all the combinations of rotations. 
In particular, the results indicate that the circular topology is the one that produces the highest entanglement, while the full topology generates the lowest entanglement, which seems contradictory since the full topology uses more entangling gates. Moreover, as the number of layers increases, the entanglement capability of the pairwise topology increases until it achieves the highest entanglement, along with the circular configuration. Furthermore, when the number of layers is $1$, the linear topology typically exhibits an entanglement value intermediate between the others. However, as the number of layers increases, its value progressively becomes closest to that of the full topology. In any case, with a high number of layers, the entanglement saturates and all topologies converge to similar values. While the circular and pairwise topologies reach an entanglement higher than $0.9$ with only $1$ or $2$ layers (except for $Ry$), the full and linear need at least $4$ layers to generate a similar level of entanglement.
Another interesting aspect is that the entanglement values depend not only on the chosen entanglement topology, but also on the individual rotations. $R_x$ rotation generates more entanglement than the $R_y$. This can be caused by the fact that the $R_x$ rotation depends on the imaginary unit, which adds a phase, while the $R_y$ does not depend on it. The single $R_z$ rotation is ignored since it only adds a phase and does not generate any entanglement.

Regarding expressibility, with $C_1$ and one layer, when considering each rotational configuration separately, the values are the same for all entanglement topologies. This suggests that expressibility may not depend on the type of entanglement but rather on the rotation used. This is confirmed when the $C_2$ circuit is considered; indeed, the results for $C_2$ with $1$ layer and $C_1$ with $2$ layers are identical, with one exception: the pairwise topology. The expressibility values of $C_2$ with $x$ layers are always identical to $C_1$ with $x+1$ layers. Even though entanglement would constrain the states that the circuit can explore, no distinguishable effect is observed. Moreover, as the number of layers increases, the circular topology exhibits the highest expressibility, followed by the pairwise topology. Nonetheless, when the number of layers increases, the expressibility converges within a range of $10^{-3} \sim 10^{-4}$ and the differences become minimal for all the entanglement topologies; however, the circular and pairwise reach this minimum with $4$ layers, while the linear and the full with $6$ layers. The single $R_z$ rotation is excluded because, although it provides expressibility similar to that of $R_x$ and $R_y$, this expressibility is due solely to phase factors and does not lead to good results. The expressibility values, with only one layer of $C_1$, are the same for $R_x$ and $R_y$. However, starting from $C_2$ and with more layers of $C_1$ or $C_2$, the expressibility values change, and the $R_x$ rotation achieves higher expressibility. This can be due to the phase factor present in $R_x$ and not in $R_y$.

\pgfplotstableread{
layer   lin     full      circ      pw
1       1.39    1.39      1.39      1.39
2       0.18    0.18      0.085     0.08
3       0.03    0.03       0.002     0.002
4       0.04    0.04      0.0002    0.0003
5       0.0006  0.0006      0.0002    0.0002
6       0.0002  0.0003      0.0002    0.0002
}\rxexp

\newcommand{\computecols}[2]{%
    \pgfplotstablecreatecol[
        create col/expr={(#2 * \thisrow{layer}) / \thisrow{lin}}
    ]{lin}{#1}

    \pgfplotstablecreatecol[
        create col/expr={(#2 * \thisrow{layer}) / \thisrow{full}}
    ]{full}{#1}

    \pgfplotstablecreatecol[
        create col/expr={(#2 * \thisrow{layer}) / \thisrow{circ}}
    ]{circ}{#1}

    \pgfplotstablecreatecol[
        create col/expr={(#2 * \thisrow{layer}) / \thisrow{pw}}
    ]{pw}{#1}
}

\pgfplotstableread{
layer   lin     full      circ      pw
1       1.39    1.39      1.39      1.39
2       0.54    0.54      0.4       0.48
3       0.31    0.31       0.208     0.24
4       0.21    0.21      0.16      0.17
5       0.18    0.18      0.16      0.16
6       0.17    0.16      0.16      0.16
}\ryexp

\pgfplotstableread{
layer   lin     full      circ      pw
1       0.72    0.72      0.72      0.72
2       0.045    0.5      0.013     0.02
3       0.005    0.009       0.0004    0.0004
4       0.0007   0.001      0.0002    0.0002
5       0.0003   0.0003      0.0002    0.0002
6       0.0002   0.0002      0.0002    0.00019
}\rxryexp
\pgfplotstableread{
layer   lin     full      circ      pw
1       0.72   0.72      0.72      0.72
2       0.068    0.09      0.031     0.05
3       0.009  0.01       0.0005    0.0005
4       0.0009  0.002      0.0002    0.00019
5       0.0003  0.0003      0.0002    0.0002
6       0.0002  0.0002      0.0002    0.0002
}\rxrzexp
\pgfplotstableread{
layer   lin     full      circ      pw
1       0.72    0.72      0.72      0.72
2       0.087    0.06      0.036     0.04
3       0.02     0.007       0.001     0.0008
4       0.002    0.0008      0.0002    0.0002
5       0.0005   0.0003      0.0002    0.00018
6       0.0003   0.0002      0.0002    0.0002
}\ryrxexp
\pgfplotstableread{
layer   lin     full      circ      pw
1       0.72    0.72      0.72      0.72
2       0.062    0.09      0.027     0.04
3       0.007    0.01       0.0005    0.0008
4       0.001    0.002      0.0002    0.00019
5       0.0003   0.0005      0.0002    0.00019
6       0.0002   0.0002      0.0002    0.0002
}\ryrzexp
\pgfplotstableread{
layer   lin     full      circ      pw
1       1.39    1.39      1.39      1.39
2       0.092    0.1      0.042     0.05
3       0.004    0.02       0.0004    0.0004
4       0.0003   0.002      0.0002    0.0002
5       0.0002    0.0004     0.0002    0.00019
6       0.0002   0.0002    0.0002    0.00019
}\rzrxexp
\pgfplotstableread{
layer   lin     full      circ      pw
1       1.39    1.39      1.39      1.39
2       0.15    0.14      0.091     0.1
3       0.02    0.02       0.002     0.003
4       0.003   0.003      0.0002    0.0002
5       0.0005  0.0006      0.0002    0.00019
6       0.0002  0.0002      0.0002    0.00019
}\rzryexp

\pgfplotstableread{
layer   lin     full      circ      pw
1       0.83    0.67      0.91      0.81
2       0.91    0.85      0.97     0.87
3       0.95    0.93       0.97     0.97
4       0.96    0.95      0.97    0.97
5       0.97    0.97      0.97    0.97
6       0.97    0.97      0.97    0.97
}\rxent

\pgfplotstableread{
layer   lin     full      circ      pw
1       0.54    0.54      0.68      0.63
2       0.64    0.64      0.8       0.844
3       0.78    0.78       0.91     0.93
4       0.82    0.82      0.92      0.93
5       0.87    0.87      0.94      0.93
6       0.89    0.89      0.94      0.93
}\ryent

\pgfplotstableread{
layer   lin     full      circ      pw
1       0.82    0.82      0.92      0.91
2       0.82    0.77      0.91     0.92
3       0.89    0.89       0.95    0.96
4       0.91    0.90      0.95    0.96
5       0.94    0.93      0.95    0.95
6       0.94    0.94      0.95    0.95
}\rxryent
\pgfplotstableread{
layer   lin     full      circ      pw
1       0.7     0.63      0.85      0.76
2       0.79    0.76      0.93     0.95
3       0.88   0.87       0.95    0.96
4       0.91   0.91      0.95    0.96
5       0.93   0.93      0.95    0.95
6       0.94   0.94      0.95    0.95
}\rxrzent
\pgfplotstableread{
layer   lin     full      circ      pw
1       0.6   0.72      0.75      0.76
2       0.73    0.8      0.87     0.92
3       0.86    0.88       0.95     0.96
4       0.89   0.91      0.95    0.96
5       0.92   0.93      0.95    0.95
6       0.94   0.94      0.95    0.95
}\ryrxent
\pgfplotstableread{
layer   lin     full      circ      pw
1       0.7    0.63      0.85      0.76
2       0.8    0.73      0.92     0.92
3       0.88   0.86       0.95    0.96
4       0.91  0.89      0.95    0.96
5       0.93  0.92      0.95    0.95
6       0.94  0.94      0.95    0.95
}\ryrzent
\pgfplotstableread{
layer   lin     full      circ      pw
1       0.83    0.55      0.91      0.82
2       0.9    0.75      0.96     0.95
3       0.94   0.86       0.96    0.96
4       0.95   0.90      0.95    0.96
5       0.96   0.92      0.95    0.95
6       0.95   0.94      0.95    0.95
}\rzrxent
\pgfplotstableread{
layer   lin     full      circ      pw
1       0.55    0.55      0.68      0.63
2       0.73    0.74      0.86      0.89
3       0.84    0.85       0.94     0.95
4       0.89    0.89      0.95    0.96
5       0.92    0.92      0.95    0.95
6       0.94    0.94      0.95    0.95
}\rzryent



\pgfplotstableread{
layer   lin     full      circ      pw
1       0.18    0.18      0.08      0.18
2       0.03    0.03      0.002     0.03
3       0.004    0.004       0.002     0.004
4       0.0006    0.0006      0.0002    0.0006
5       0.0003  0.0003      0.0002    0.0003
6       0.0002  0.0003      0.0002    0.0002
}\rxexpc

\pgfplotstableread{
layer   lin     full      circ      pw
1       0.54    0.54      0.41      0.61
2       0.3    0.35      0.2       0.35
3       0.21    0.21       0.16     0.22
4       0.18    0.18      0.16      0.17
5       0.16    0.17      0.16      0.16
6       0.16    0.16      0.16      0.16
}\ryexpc

\pgfplotstableread{
layer   lin     full      circ      pw
1       0.05    0.05      0.01      0.06
2       0.004    0.01      0.0003     0.007
3       0.0007    0.001       0.0002    0.0007
4       0.0003   0.0004      0.0002    0.0002
5       0.0002   0.0002      0.0002    0.0002
6       0.0002   0.0002      0.0002    0.00019
}\rxryexpc
\pgfplotstableread{
layer   lin     full      circ      pw
1       0.07   0.09      0.03      0.1
2       0.08    0.01      0.0005     0.01
3       0.001  0.001       0.0002    0.001
4       0.0003  0.0003      0.0002    0.00019
5       0.0002  0.0002      0.0002    0.0002
6       0.0002  0.0002      0.0002    0.0002
}\rxrzexpc
\pgfplotstableread{
layer   lin     full      circ      pw
1       0.09    0.06      0.04      0.09
2       0.1    0.006      0.001     0.01
3       0.02     0.0008       0.0002     0.001
4       0.0005    0.0003      0.0002    0.0003
5       0.0002   0.0002      0.0002    0.00018
6       0.0002   0.0002      0.0002    0.0002
}\ryrxexpc
\pgfplotstableread{
layer   lin     full      circ      pw
1       0.06    0.09      0.03      0.09
2       0.07    0.01      0.0005     0.01
3       0.002    0.002       0.0002    0.001
4       0.0003    0.0005      0.0002    0.0003
5       0.0002   0.0002      0.0002    0.00019
6       0.0002   0.0002      0.0002    0.0002
}\ryrzexpc
\pgfplotstableread{
layer   lin     full      circ      pw
1       0.09    0.1      0.04      0.11
2       0.04    0.015      0.0004     0.007
3       0.0009    0.002       0.0002    0.0006
4       0.0002   0.0004      0.0002    0.0002
5       0.0002    0.0002     0.0002    0.00019
6       0.0002   0.0002    0.0002    0.00019
}\rzrxexpc
\pgfplotstableread{
layer   lin     full      circ      pw
1       0.15    0.13      0.09      0.17
2       0.02    0.02      0.002     0.02
3       0.003    0.003       0.0002     0.003
4       0.0006   0.0005      0.0002    0.0004
5       0.0003  0.0002      0.0002    0.00019
6       0.0002  0.0002      0.0002    0.00019
}\rzryexpc



\pgfplotstableread{
layer   lin     full      circ      pw
1       0.83    0.67      0.91      0.81
2       0.91    0.85      0.97     0.87
3       0.95    0.93       0.97     0.97
4       0.96    0.95      0.97    0.97
5       0.97    0.97      0.97    0.97
6       0.97    0.97      0.97    0.97
}\rxentc

\pgfplotstableread{
layer   lin     full      circ      pw
1       0.54    0.54      0.68      0.63
2       0.64    0.64      0.8       0.844
3       0.78    0.78       0.91     0.93
4       0.82    0.82      0.92      0.93
5       0.87    0.87      0.94      0.93
6       0.89    0.89      0.94      0.93
}\ryentc

\pgfplotstableread{
layer   lin     full      circ      pw
1       0.82    0.82      0.92      0.91
2       0.82    0.77      0.91     0.92
3       0.89    0.89       0.95    0.96
4       0.91    0.90      0.95    0.96
5       0.94    0.93      0.95    0.95
6       0.94    0.94      0.95    0.95
}\rxryentc
\pgfplotstableread{
layer   lin     full      circ      pw
1       0.7     0.63      0.85      0.76
2       0.79    0.76      0.93     0.95
3       0.88   0.87       0.95    0.96
4       0.91   0.91      0.95    0.96
5       0.93   0.93      0.95    0.95
6       0.94   0.94      0.95    0.95
}\rxrzentc
\pgfplotstableread{
layer   lin     full      circ      pw
1       0.6   0.72      0.75      0.76
2       0.73    0.8      0.87     0.92
3       0.86    0.88       0.95     0.96
4       0.89   0.91      0.95    0.96
5       0.92   0.93      0.95    0.95
6       0.94   0.94      0.95    0.95
}\ryrxentc
\pgfplotstableread{
layer   lin     full      circ      pw
1       0.7    0.63      0.85      0.76
2       0.8    0.73      0.92     0.92
3       0.88   0.86       0.95    0.96
4       0.91  0.89      0.95    0.96
5       0.93  0.92      0.95    0.95
6       0.94  0.94      0.95    0.95
}\ryrzentc
\pgfplotstableread{
layer   lin     full      circ      pw
1       0.83    0.55      0.91      0.82
2       0.9    0.75      0.96     0.95
3       0.94   0.86       0.96    0.96
4       0.95   0.90      0.95    0.96
5       0.96   0.92      0.95    0.95
6       0.95   0.94      0.95    0.95
}\rzrxentc
\pgfplotstableread{
layer   lin     full      circ      pw
1       0.55    0.55      0.68      0.63
2       0.73    0.74      0.86      0.89
3       0.84    0.85       0.94     0.95
4       0.89    0.89      0.95    0.96
5       0.92    0.92      0.95    0.95
6       0.94    0.94      0.95    0.95
}\rzryentc

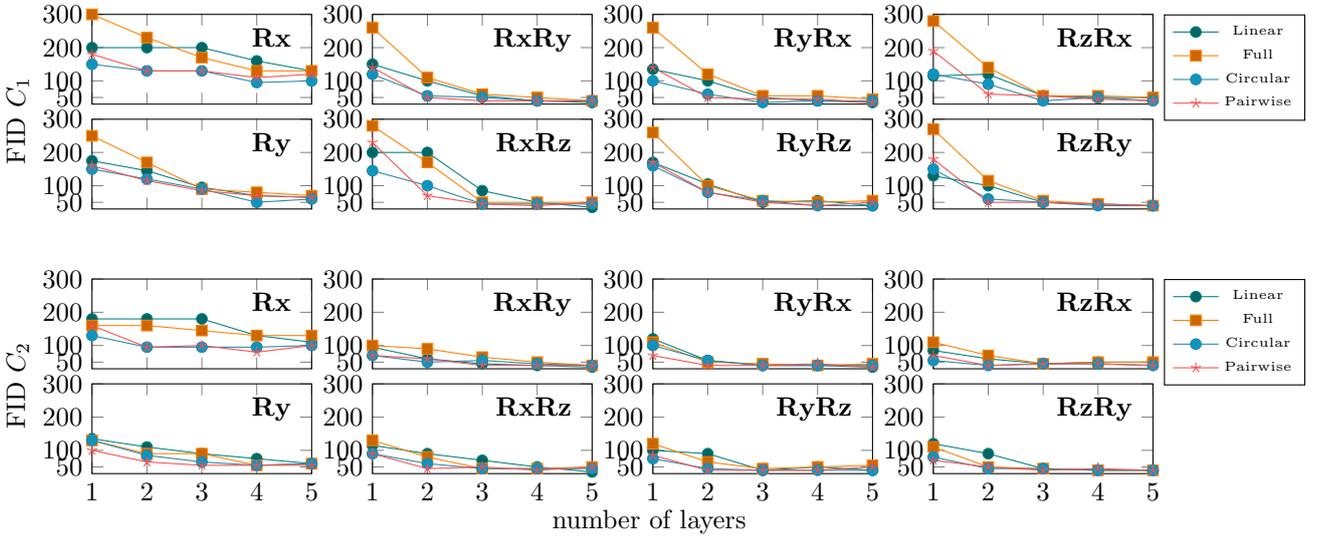
\begin{figure*}[htb]
    \centering
    \begin{subfigure}{\textwidth}
        \centering
        \begin{tikzpicture}
            \begin{groupplot}[
                group style={
                    group size=4 by 2,
                    vertical sep=0.2cm,
                    horizontal sep=0.8cm,  
                    group name=myplots2,
                },
                width=0.17\columnwidth,    
                height=0.07\columnwidth,
                scale only axis,
                title={},
                xmin=1, xmax=5,
                xtick={1, 2, 3, 4, 5},
                samples at={1, 2, 3, 4, 5},
                ymin=30, ymax=300,
                ytick = {50, 100, 200, 300},
                cycle list name=exotic,
                legend style={font=\tiny, at={(1.05,1)}, anchor=north west},
            ]

            \nextgroupplot[xticklabels=\empty] 
            \addplot table[x=layer, y=lin] {\rxfid};
            \addplot table[x=layer, y=full] {\rxfid};
            \addplot table[x=layer, y=circ] {\rxfid};
            \addplot table[x=layer, y=pw] {\rxfid};
            \node[anchor=north east, font=\bfseries] at (rel axis cs:0.95, 0.95) {Rx};

        \nextgroupplot[xticklabels=\empty] 
            \addplot table[x=layer, y=lin] {\rxryfid};
            \addplot table[x=layer, y=full] {\rxryfid};
            \addplot table[x=layer, y=circ] {\rxryfid};
            \addplot table[x=layer, y=pw] {\rxryfid};
            \node[anchor=north east, font=\bfseries] at (rel axis cs:0.95, 0.95) {RxRy};

        \nextgroupplot[xticklabels=\empty] 
            \addplot table[x=layer, y=lin] {\ryrxfid};
            \addplot table[x=layer, y=full] {\ryrxfid};
            \addplot table[x=layer, y=circ] {\ryrxfid};
            \addplot table[x=layer, y=pw] {\ryrxfid};
            \node[anchor=north east, font=\bfseries] at (rel axis cs:0.95, 0.95) {RyRx};
            
        \nextgroupplot[xticklabels=\empty] 
            \addplot table[x=layer, y=lin] {\rzrxfid}; \addlegendentry{Linear}
            \addplot table[x=layer, y=full] {\rzrxfid}; \addlegendentry{Full}
            \addplot table[x=layer, y=circ] {\rzrxfid}; \addlegendentry{Circular}
            \addplot table[x=layer, y=pw] {\rzrxfid}; \addlegendentry{Pairwise}
            \node[anchor=north east, font=\bfseries] at (rel axis cs:0.95, 0.95) {RzRx};
        

        \nextgroupplot[xticklabels=\empty] 
            \addplot table[x=layer, y=lin] {\ryfid};
            \addplot table[x=layer, y=full] {\ryfid};
            \addplot table[x=layer, y=circ] {\ryfid};
            \addplot table[x=layer, y=pw] {\ryfid};
            \node[anchor=north east, font=\bfseries] at (rel axis cs:0.95, 0.95) {Ry};

        \nextgroupplot[xticklabels=\empty] 
            \addplot table[x=layer, y=lin] {\rxrzfid};
            \addplot table[x=layer, y=full] {\rxrzfid};
            \addplot table[x=layer, y=circ] {\rxrzfid};
            \addplot table[x=layer, y=pw] {\rxrzfid};
            \node[anchor=north east, font=\bfseries] at (rel axis cs:0.95, 0.95) {RxRz};

        \nextgroupplot[xticklabels=\empty] 
            \addplot table[x=layer, y=lin] {\ryrzfid};
            \addplot table[x=layer, y=full] {\ryrzfid};
            \addplot table[x=layer, y=circ] {\ryrzfid};
            \addplot table[x=layer, y=pw] {\ryrzfid};
            \node[anchor=north east, font=\bfseries] at (rel axis cs:0.95, 0.95) {RyRz};
            
        \nextgroupplot[xticklabels=\empty] 
            \addplot table[x=layer, y=lin] {\rzryfid};
            \addplot table[x=layer, y=full] {\rzryfid};
            \addplot table[x=layer, y=circ] {\rzryfid};
            \addplot table[x=layer, y=pw] {\rzryfid};
            \node[anchor=north east, font=\bfseries] at (rel axis cs:0.95, 0.95) {RzRy};
            
        \end{groupplot}

            \node at (myplots c1r1.north west) [rotate=90, anchor=south, yshift=2em, xshift=-4em] {FID $C_1$};

        \end{tikzpicture}
        \label{fig:fid_c1_sub}
    \end{subfigure}


    \begin{subfigure}{\textwidth}
        \centering
        \begin{tikzpicture}
            \begin{groupplot}[
                group style={
                    group size=4 by 2,
                    vertical sep=0.2cm,
                    horizontal sep=0.8cm,  
                    group name=myplots2,
                },
                width=0.17\columnwidth,    
                height=0.07\columnwidth,
                scale only axis,
                title={},
                xmin=1, xmax=5,
                xtick={1, 2, 3, 4, 5},
                samples at={1, 2, 3, 4, 5},
                ymin=30, ymax=300,
                ytick = {50, 100, 200, 300},
                cycle list name=exotic,
                legend style={font=\tiny, at={(1.05,1)}, anchor=north west},
            ]

        
        \nextgroupplot[xticklabels=\empty] 
            \addplot table[x=layer, y=lin] {\rxfidc};
            \addplot table[x=layer, y=full] {\rxfidc};
            \addplot table[x=layer, y=circ] {\rxfidc};
            \addplot table[x=layer, y=pw] {\rxfidc};
            \node[anchor=north east, font=\bfseries] at (rel axis cs:0.95, 0.95) {Rx};

        \nextgroupplot[xticklabels=\empty] 
            \addplot table[x=layer, y=lin] {\rxryfidc};
            \addplot table[x=layer, y=full] {\rxryfidc};
            \addplot table[x=layer, y=circ] {\rxryfidc};
            \addplot table[x=layer, y=pw] {\rxryfidc};
            \node[anchor=north east, font=\bfseries] at (rel axis cs:0.95, 0.95) {RxRy};

        \nextgroupplot[xticklabels=\empty] 
            \addplot table[x=layer, y=lin] {\ryrxfidc};
            \addplot table[x=layer, y=full] {\ryrxfidc};
            \addplot table[x=layer, y=circ] {\ryrxfidc};
            \addplot table[x=layer, y=pw] {\ryrxfidc};
            \node[anchor=north east, font=\bfseries] at (rel axis cs:0.95, 0.95) {RyRx};
            
        \nextgroupplot[xticklabels=\empty] 
            \addplot table[x=layer, y=lin] {\rzrxfidc}; \addlegendentry{Linear}
            \addplot table[x=layer, y=full] {\rzrxfidc}; \addlegendentry{Full}
            \addplot table[x=layer, y=circ] {\rzrxfidc}; \addlegendentry{Circular}
            \addplot table[x=layer, y=pw] {\rzrxfidc}; \addlegendentry{Pairwise}
            \node[anchor=north east, font=\bfseries] at (rel axis cs:0.95, 0.95) {RzRx};
        

        \nextgroupplot 
            \addplot table[x=layer, y=lin] {\ryfidc};
            \addplot table[x=layer, y=full] {\ryfidc};
            \addplot table[x=layer, y=circ] {\ryfidc};
            \addplot table[x=layer, y=pw] {\ryfidc};
            \node[anchor=north east, font=\bfseries] at (rel axis cs:0.95, 0.95) {Ry};

        \nextgroupplot 
            \addplot table[x=layer, y=lin] {\rxrzfidc};
            \addplot table[x=layer, y=full] {\rxrzfidc};
            \addplot table[x=layer, y=circ] {\rxrzfidc};
            \addplot table[x=layer, y=pw] {\rxrzfidc};
            \node[anchor=north east, font=\bfseries] at (rel axis cs:0.95, 0.95) {RxRz};

        \nextgroupplot 
            \addplot table[x=layer, y=lin] {\ryrzfidc};
            \addplot table[x=layer, y=full] {\ryrzfidc};
            \addplot table[x=layer, y=circ] {\ryrzfidc};
            \addplot table[x=layer, y=pw] {\ryrzfidc};
            \node[anchor=north east, font=\bfseries] at (rel axis cs:0.95, 0.95) {RyRz};
            
        \nextgroupplot 
            \addplot table[x=layer, y=lin] {\rzryfidc};
            \addplot table[x=layer, y=full] {\rzryfidc};
            \addplot table[x=layer, y=circ] {\rzryfidc};
            \addplot table[x=layer, y=pw] {\rzryfidc};
            \node[anchor=north east, font=\bfseries] at (rel axis cs:0.95, 0.95) {RzRy};
            
        \end{groupplot}

            \node[anchor=north] at ($(myplots2 c2r2.south)!0.5!(myplots2 c3r2.south)$) [yshift=-1em, xshift=1em] {number of layers};
            \node at (myplots2 c1r1.north west) [rotate=90, anchor=south, yshift=2em, xshift=-4em] {FID $C_2$};

        \end{tikzpicture}
        \label{fig:fid_c2_sub}
    \end{subfigure}
    \captionsetup{skip=-0.05pt}
    \caption{FID results of $C_1$ and $C_2$.}
    \label{fig:combined_fid}
\end{figure*}

\begin{figure*}[htb]
    \centering
    \begin{subfigure}{\textwidth}
        \centering
        \begin{tikzpicture}
            \begin{groupplot}[
                group style={
                    group size=4 by 2,
                    vertical sep=0.2cm,
                    horizontal sep=0.8cm,  
                    group name=myplots2,
                },
                width=0.17\columnwidth,    
                height=0.07\columnwidth,
                scale only axis,
                title={},
                xmin=1, xmax=5,
                xtick={1, 2, 3, 4, 5},
                samples at={1, 2, 3, 4, 5},
                ymin=0.25, ymax=0.55,
                ytick = {0.3, 0.4, 0.5},
                cycle list name=exotic,
                legend style={font=\tiny, at={(1.05,1)}, anchor=north west},
            ]

            \nextgroupplot[xticklabels=\empty] 
            \addplot table[x=layer, y=lin] {\rxhell};
            \addplot table[x=layer, y=full] {\rxhell};
            \addplot table[x=layer, y=circ] {\rxhell};
            \addplot table[x=layer, y=pw] {\rxhell};
            \node[anchor=north east, font=\bfseries] at (rel axis cs:0.95, 0.95) {Rx};

        \nextgroupplot[xticklabels=\empty] 
            \addplot table[x=layer, y=lin] {\rxryhell};
            \addplot table[x=layer, y=full] {\rxryhell};
            \addplot table[x=layer, y=circ] {\rxryhell};
            \addplot table[x=layer, y=pw] {\rxryhell};
            \node[anchor=north east, font=\bfseries] at (rel axis cs:0.95, 0.95) {RxRy};

        \nextgroupplot[xticklabels=\empty] 
            \addplot table[x=layer, y=lin] {\ryrxhell};
            \addplot table[x=layer, y=full] {\ryrxhell};
            \addplot table[x=layer, y=circ] {\ryrxhell};
            \addplot table[x=layer, y=pw] {\ryrxhell};
            \node[anchor=north east, font=\bfseries] at (rel axis cs:0.95, 0.95) {RyRx};
            
        \nextgroupplot[xticklabels=\empty] 
            \addplot table[x=layer, y=lin] {\rzrxhell}; \addlegendentry{Linear}
            \addplot table[x=layer, y=full] {\rzrxhell}; \addlegendentry{Full}
            \addplot table[x=layer, y=circ] {\rzrxhell}; \addlegendentry{Circular}
            \addplot table[x=layer, y=pw] {\rzrxhell}; \addlegendentry{Pairwise}
            \node[anchor=north east, font=\bfseries] at (rel axis cs:0.95, 0.95) {RzRx};
        

        \nextgroupplot[xticklabels=\empty] 
            \addplot table[x=layer, y=lin] {\ryhell};
            \addplot table[x=layer, y=full] {\ryhell};
            \addplot table[x=layer, y=circ] {\ryhell};
            \addplot table[x=layer, y=pw] {\ryhell};
            \node[anchor=north east, font=\bfseries] at (rel axis cs:0.95, 0.95) {Ry};

        \nextgroupplot[xticklabels=\empty] 
            \addplot table[x=layer, y=lin] {\rxrzhell};
            \addplot table[x=layer, y=full] {\rxrzhell};
            \addplot table[x=layer, y=circ] {\rxrzhell};
            \addplot table[x=layer, y=pw] {\rxrzhell};
            \node[anchor=north east, font=\bfseries] at (rel axis cs:0.95, 0.95) {RxRz};

        \nextgroupplot[xticklabels=\empty] 
            \addplot table[x=layer, y=lin] {\ryrzhell};
            \addplot table[x=layer, y=full] {\ryrzhell};
            \addplot table[x=layer, y=circ] {\ryrzhell};
            \addplot table[x=layer, y=pw] {\ryrzhell};
            \node[anchor=north east, font=\bfseries] at (rel axis cs:0.95, 0.95) {RyRz};
            
        \nextgroupplot[xticklabels=\empty] 
            \addplot table[x=layer, y=lin] {\rzryhell};
            \addplot table[x=layer, y=full] {\rzryhell};
            \addplot table[x=layer, y=circ] {\rzryhell};
            \addplot table[x=layer, y=pw] {\rzryhell};
            \node[anchor=north east, font=\bfseries] at (rel axis cs:0.95, 0.95) {RzRy};
            
        \end{groupplot}

            \node at (myplots c1r1.north west) [rotate=90, anchor=south, yshift=2em, xshift=-4em] {Hellinger $C_1$};

        \end{tikzpicture}
        \label{fig:hell_c1_sub}
    \end{subfigure}


    \begin{subfigure}{\textwidth}
        \centering
        \begin{tikzpicture}
            \begin{groupplot}[
                group style={
                    group size=4 by 2,
                    vertical sep=0.2cm,
                    horizontal sep=0.8cm,  
                    group name=myplots2,
                },
                width=0.17\columnwidth,    
                height=0.07\columnwidth,
                scale only axis,
                title={},
                xmin=1, xmax=5,
                xtick={1, 2, 3, 4, 5},
                samples at={1, 2, 3, 4, 5},
                ymin=0.25, ymax=0.55,
                ytick = {0.3, 0.4, 0.5},
                cycle list name=exotic,
                legend style={font=\tiny, at={(1.05,1)}, anchor=north west},
            ]

        
        \nextgroupplot[xticklabels=\empty] 
            \addplot table[x=layer, y=lin] {\rxhellc};
            \addplot table[x=layer, y=full] {\rxhellc};
            \addplot table[x=layer, y=circ] {\rxhellc};
            \addplot table[x=layer, y=pw] {\rxhellc};
            \node[anchor=north east, font=\bfseries] at (rel axis cs:0.95, 0.95) {Rx};

        \nextgroupplot[xticklabels=\empty] 
            \addplot table[x=layer, y=lin] {\rxryhellc};
            \addplot table[x=layer, y=full] {\rxryhellc};
            \addplot table[x=layer, y=circ] {\rxryhellc};
            \addplot table[x=layer, y=pw] {\rxryhellc};
            \node[anchor=north east, font=\bfseries] at (rel axis cs:0.95, 0.95) {RxRy};

        \nextgroupplot[xticklabels=\empty] 
            \addplot table[x=layer, y=lin] {\ryrxhellc};
            \addplot table[x=layer, y=full] {\ryrxhellc};
            \addplot table[x=layer, y=circ] {\ryrxhellc};
            \addplot table[x=layer, y=pw] {\ryrxhellc};
            \node[anchor=north east, font=\bfseries] at (rel axis cs:0.95, 0.95) {RyRx};
            
        \nextgroupplot[xticklabels=\empty] 
            \addplot table[x=layer, y=lin] {\rzrxhellc}; \addlegendentry{Linear}
            \addplot table[x=layer, y=full] {\rzrxhellc}; \addlegendentry{Full}
            \addplot table[x=layer, y=circ] {\rzrxhellc}; \addlegendentry{Circular}
            \addplot table[x=layer, y=pw] {\rzrxhellc}; \addlegendentry{Pairwise}
            \node[anchor=north east, font=\bfseries] at (rel axis cs:0.95, 0.95) {RzRx};
        

        \nextgroupplot 
            \addplot table[x=layer, y=lin] {\ryhellc};
            \addplot table[x=layer, y=full] {\ryhellc};
            \addplot table[x=layer, y=circ] {\ryhellc};
            \addplot table[x=layer, y=pw] {\ryhellc};
            \node[anchor=north east, font=\bfseries] at (rel axis cs:0.95, 0.95) {Ry};

        \nextgroupplot 
            \addplot table[x=layer, y=lin] {\rxrzhellc};
            \addplot table[x=layer, y=full] {\rxrzhellc};
            \addplot table[x=layer, y=circ] {\rxrzhellc};
            \addplot table[x=layer, y=pw] {\rxrzhellc};
            \node[anchor=north east, font=\bfseries] at (rel axis cs:0.95, 0.95) {RxRz};

        \nextgroupplot 
            \addplot table[x=layer, y=lin] {\ryrzhellc};
            \addplot table[x=layer, y=full] {\ryrzhellc};
            \addplot table[x=layer, y=circ] {\ryrzhellc};
            \addplot table[x=layer, y=pw] {\ryrzhellc};
            \node[anchor=north east, font=\bfseries] at (rel axis cs:0.95, 0.95) {RyRz};
            
        \nextgroupplot 
            \addplot table[x=layer, y=lin] {\rzryhellc};
            \addplot table[x=layer, y=full] {\rzryhellc};
            \addplot table[x=layer, y=circ] {\rzryhellc};
            \addplot table[x=layer, y=pw] {\rzryhellc};
            \node[anchor=north east, font=\bfseries] at (rel axis cs:0.95, 0.95) {RzRy};
            
        \end{groupplot}

            \node[anchor=north] at ($(myplots2 c2r2.south)!0.5!(myplots2 c3r2.south)$) [yshift=-1em, xshift=1em] {number of layers};
            \node at (myplots2 c1r1.north west) [rotate=90, anchor=south, yshift=2em, xshift=-4em] {Hellinger $C_2$};

        \end{tikzpicture}
        \label{fig:hell_c2_sub}
    \end{subfigure}
    \captionsetup{skip=-0.05pt}
    \caption{Hellinger distance results of $C_1$ and $C_2$.}
    \label{fig:combined_hell}
\end{figure*}
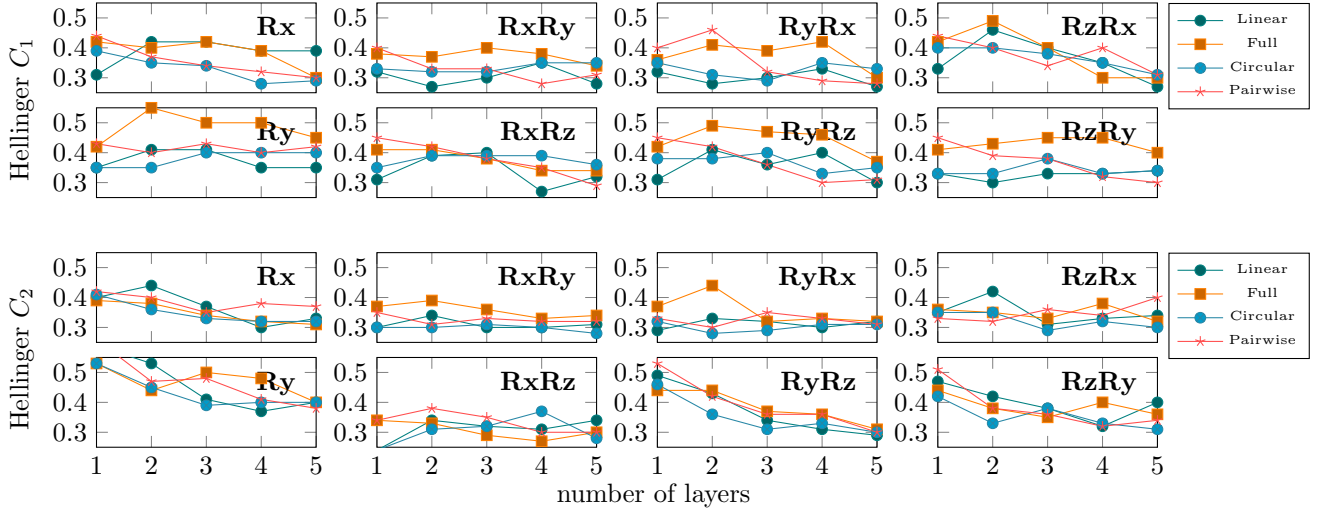

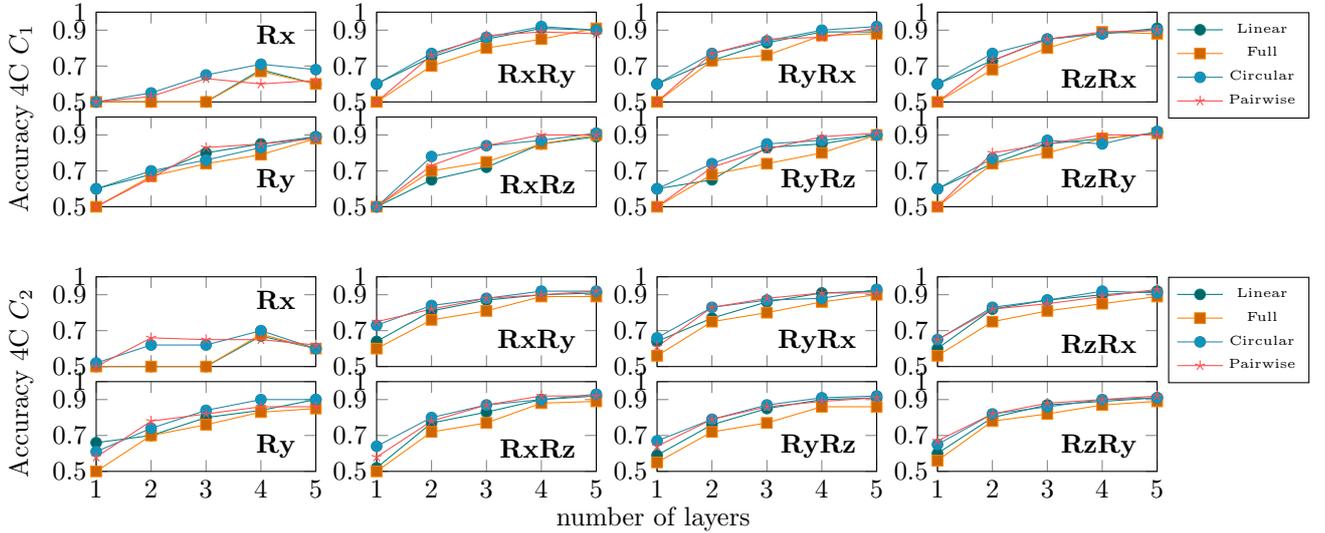
\begin{figure*}[htb]
    \centering
    \begin{subfigure}{\textwidth}
        \centering
        \begin{tikzpicture}
            \begin{groupplot}[
                group style={
                    group size=4 by 2,
                    vertical sep=0.2cm,
                    horizontal sep=0.8cm,  
                    group name=myplots2,
                },
                width=0.17\columnwidth,    
                height=0.07\columnwidth,
                scale only axis,
                title={},
                xmin=1, xmax=5,
                xtick={1, 2, 3, 4, 5},
                samples at={1, 2, 3, 4, 5},
                ymin=0.5, ymax=1,
                ytick={0.5, 0.7, 0.9, 1},
                cycle list name=exotic,
                legend style={font=\tiny, at={(1.05,1)}, anchor=north west},
            ]

            \nextgroupplot[xticklabels=\empty] 
            \addplot table[x=layer, y=lin] {\rxacc};
            \addplot table[x=layer, y=full] {\rxacc};
            \addplot table[x=layer, y=circ] {\rxacc};
            \addplot table[x=layer, y=pw] {\rxacc};
            \node[anchor=north east, font=\bfseries] at (rel axis cs:0.95, 0.95) {Rx};

        \nextgroupplot[xticklabels=\empty] 
            \addplot table[x=layer, y=lin] {\rxryacc};
            \addplot table[x=layer, y=full] {\rxryacc};
            \addplot table[x=layer, y=circ] {\rxryacc};
            \addplot table[x=layer, y=pw] {\rxryacc};
            \node[anchor=south east, font=\bfseries] at (rel axis cs:0.95, 0.05) {RxRy};

        \nextgroupplot[xticklabels=\empty] 
            \addplot table[x=layer, y=lin] {\ryrxacc};
            \addplot table[x=layer, y=full] {\ryrxacc};
            \addplot table[x=layer, y=circ] {\ryrxacc};
            \addplot table[x=layer, y=pw] {\ryrxacc};
            \node[anchor=south east, font=\bfseries] at (rel axis cs:0.95, 0.05) {RyRx};
            
        \nextgroupplot[xticklabels=\empty] 
            \addplot table[x=layer, y=lin] {\rzrxacc}; \addlegendentry{Linear}
            \addplot table[x=layer, y=full] {\rzrxacc}; \addlegendentry{Full}
            \addplot table[x=layer, y=circ] {\rzrxacc}; \addlegendentry{Circular}
            \addplot table[x=layer, y=pw] {\rzrxacc}; \addlegendentry{Pairwise}
            \node[anchor=south east, font=\bfseries] at (rel axis cs:0.95, 0.05) {RzRx};
        

        \nextgroupplot[xticklabels=\empty] 
            \addplot table[x=layer, y=lin] {\ryacc};
            \addplot table[x=layer, y=full] {\ryacc};
            \addplot table[x=layer, y=circ] {\ryacc};
            \addplot table[x=layer, y=pw] {\ryacc};
            \node[anchor=south east, font=\bfseries] at (rel axis cs:0.95, 0.05) {Ry};

        \nextgroupplot[xticklabels=\empty] 
            \addplot table[x=layer, y=lin] {\rxrzacc};
            \addplot table[x=layer, y=full] {\rxrzacc};
            \addplot table[x=layer, y=circ] {\rxrzacc};
            \addplot table[x=layer, y=pw] {\rxrzacc};
            \node[anchor=south east, font=\bfseries] at (rel axis cs:0.95, 0.05) {RxRz};

        \nextgroupplot[xticklabels=\empty] 
            \addplot table[x=layer, y=lin] {\ryrzacc};
            \addplot table[x=layer, y=full] {\ryrzacc};
            \addplot table[x=layer, y=circ] {\ryrzacc};
            \addplot table[x=layer, y=pw] {\ryrzacc};
            \node[anchor=south east, font=\bfseries] at (rel axis cs:0.95, 0.05) {RyRz};
            
        \nextgroupplot[xticklabels=\empty] 
            \addplot table[x=layer, y=lin] {\rzryacc};
            \addplot table[x=layer, y=full] {\rzryacc};
            \addplot table[x=layer, y=circ] {\rzryacc};
            \addplot table[x=layer, y=pw] {\rzryacc};
            \node[anchor=south east, font=\bfseries] at (rel axis cs:0.95, 0.05) {RzRy};
            
        \end{groupplot}

            \node at (myplots c1r1.north west) [rotate=90, anchor=south, yshift=2em, xshift=-4em] {Accuracy $4$C $C_1$};

        \end{tikzpicture}
        \label{fig:class4_c1_sub}
    \end{subfigure}


    \begin{subfigure}{\textwidth}
        \centering
        \begin{tikzpicture}
            \begin{groupplot}[
                group style={
                    group size=4 by 2,
                    vertical sep=0.2cm,
                    horizontal sep=0.8cm,  
                    group name=myplots2,
                },
                width=0.17\columnwidth,    
                height=0.07\columnwidth,
                scale only axis,
                title={},
                xmin=1, xmax=5,
                xtick={1, 2, 3, 4, 5},
                samples at={1, 2, 3, 4, 5},
                ymin=0.5, ymax=1,
                ytick={0.5, 0.7, 0.9, 1},
                cycle list name=exotic,
                legend style={font=\tiny, at={(1.05,1)}, anchor=north west},
            ]

        
        \nextgroupplot[xticklabels=\empty] 
            \addplot table[x=layer, y=lin] {\rxaccx};
            \addplot table[x=layer, y=full] {\rxaccx};
            \addplot table[x=layer, y=circ] {\rxaccx};
            \addplot table[x=layer, y=pw] {\rxaccx};
            \node[anchor=north east, font=\bfseries] at (rel axis cs:0.95, 0.95) {Rx};

        \nextgroupplot[xticklabels=\empty] 
            \addplot table[x=layer, y=lin] {\rxryaccx};
            \addplot table[x=layer, y=full] {\rxryaccx};
            \addplot table[x=layer, y=circ] {\rxryaccx};
            \addplot table[x=layer, y=pw] {\rxryaccx};
            \node[anchor=south east, font=\bfseries] at (rel axis cs:0.95, 0.05) {RxRy};

        \nextgroupplot[xticklabels=\empty] 
            \addplot table[x=layer, y=lin] {\ryrxaccx};
            \addplot table[x=layer, y=full] {\ryrxaccx};
            \addplot table[x=layer, y=circ] {\ryrxaccx};
            \addplot table[x=layer, y=pw] {\ryrxaccx};
            \node[anchor=south east, font=\bfseries] at (rel axis cs:0.95, 0.05) {RyRx};
            
        \nextgroupplot[xticklabels=\empty] 
            \addplot table[x=layer, y=lin] {\rzrxaccx}; \addlegendentry{Linear}
            \addplot table[x=layer, y=full] {\rzrxaccx}; \addlegendentry{Full}
            \addplot table[x=layer, y=circ] {\rzrxaccx}; \addlegendentry{Circular}
            \addplot table[x=layer, y=pw] {\rzrxaccx}; \addlegendentry{Pairwise}
            \node[anchor=south east, font=\bfseries] at (rel axis cs:0.95, 0.05) {RzRx};
        

        \nextgroupplot 
            \addplot table[x=layer, y=lin] {\ryaccx};
            \addplot table[x=layer, y=full] {\ryaccx};
            \addplot table[x=layer, y=circ] {\ryaccx};
            \addplot table[x=layer, y=pw] {\ryaccx};
            \node[anchor=south east, font=\bfseries] at (rel axis cs:0.95, 0.05) {Ry};

        \nextgroupplot 
            \addplot table[x=layer, y=lin] {\rxrzaccx};
            \addplot table[x=layer, y=full] {\rxrzaccx};
            \addplot table[x=layer, y=circ] {\rxrzaccx};
            \addplot table[x=layer, y=pw] {\rxrzaccx};
            \node[anchor=south east, font=\bfseries] at (rel axis cs:0.95, 0.05) {RxRz};

        \nextgroupplot 
            \addplot table[x=layer, y=lin] {\ryrzaccx};
            \addplot table[x=layer, y=full] {\ryrzaccx};
            \addplot table[x=layer, y=circ] {\ryrzaccx};
            \addplot table[x=layer, y=pw] {\ryrzaccx};
            \node[anchor=south east, font=\bfseries] at (rel axis cs:0.95, 0.05) {RyRz};
            
        \nextgroupplot 
            \addplot table[x=layer, y=lin] {\rzryaccx};
            \addplot table[x=layer, y=full] {\rzryaccx};
            \addplot table[x=layer, y=circ] {\rzryaccx};
            \addplot table[x=layer, y=pw] {\rzryaccx};
            \node[anchor=south east, font=\bfseries] at (rel axis cs:0.95, 0.05) {RzRy};
            
        \end{groupplot}

            \node[anchor=north] at ($(myplots2 c2r2.south)!0.5!(myplots2 c3r2.south)$) [yshift=-1em, xshift=1em] {number of layers};
            \node at (myplots2 c1r1.north west) [rotate=90, anchor=south, yshift=2em, xshift=-4em] {Accuracy $4$C $C_2$};

        \end{tikzpicture}
        \label{fig:class6_c1_sub}
    \end{subfigure}
    \captionsetup{skip=-0.05pt}
    \caption{Classification accuracy results of $C_1$ and $C_2$ with $4$ classes.}
    \label{fig:combined_class4}
\end{figure*}

\begin{figure*}[htb]
    \centering
    \begin{subfigure}{\textwidth}
        \centering
        \begin{tikzpicture}
            \begin{groupplot}[
                group style={
                    group size=4 by 2,
                    vertical sep=0.2cm,
                    horizontal sep=0.8cm,  
                    group name=myplots2,
                },
                width=0.17\columnwidth,    
                height=0.07\columnwidth,
                scale only axis,
                title={},
                xmin=1, xmax=5,
                xtick={1, 2, 3, 4, 5},
                samples at={1, 2, 3, 4, 5},
                ymin=0.3, ymax=1,
                ytick={0.3, 0.5, 0.7, 0.9},
                cycle list name=exotic,
                legend style={font=\tiny, at={(1.05,1)}, anchor=north west},
            ]

            \nextgroupplot[xticklabels=\empty] 
            \addplot table[x=layer, y=lin] {\rxaccc};
            \addplot table[x=layer, y=full] {\rxaccc};
            \addplot table[x=layer, y=circ] {\rxaccc};
            \addplot table[x=layer, y=pw] {\rxaccc};
            \node[anchor=north east, font=\bfseries] at (rel axis cs:0.95, 0.95) {Rx};

        \nextgroupplot[xticklabels=\empty] 
            \addplot table[x=layer, y=lin] {\rxryaccc};
            \addplot table[x=layer, y=full] {\rxryaccc};
            \addplot table[x=layer, y=circ] {\rxryaccc};
            \addplot table[x=layer, y=pw] {\rxryaccc};
            \node[anchor=south east, font=\bfseries] at (rel axis cs:0.95, 0.05) {RxRy};

        \nextgroupplot[xticklabels=\empty] 
            \addplot table[x=layer, y=lin] {\ryrxaccc};
            \addplot table[x=layer, y=full] {\ryrxaccc};
            \addplot table[x=layer, y=circ] {\ryrxaccc};
            \addplot table[x=layer, y=pw] {\ryrxaccc};
            \node[anchor=south east, font=\bfseries] at (rel axis cs:0.95, 0.05) {RyRx};
            
        \nextgroupplot[xticklabels=\empty] 
            \addplot table[x=layer, y=lin] {\rzrxaccc}; \addlegendentry{Linear}
            \addplot table[x=layer, y=full] {\rzrxaccc}; \addlegendentry{Full}
            \addplot table[x=layer, y=circ] {\rzrxaccc}; \addlegendentry{Circular}
            \addplot table[x=layer, y=pw] {\rzrxaccc}; \addlegendentry{Pairwise}
            \node[anchor=south east, font=\bfseries] at (rel axis cs:0.95, 0.05) {RzRx};
        

        \nextgroupplot[xticklabels=\empty] 
            \addplot table[x=layer, y=lin] {\ryaccc};
            \addplot table[x=layer, y=full] {\ryaccc};
            \addplot table[x=layer, y=circ] {\ryaccc};
            \addplot table[x=layer, y=pw] {\ryaccc};
            \node[anchor=south east, font=\bfseries] at (rel axis cs:0.95, 0.05) {Ry};

        \nextgroupplot[xticklabels=\empty] 
            \addplot table[x=layer, y=lin] {\rxrzaccc};
            \addplot table[x=layer, y=full] {\rxrzaccc};
            \addplot table[x=layer, y=circ] {\rxrzaccc};
            \addplot table[x=layer, y=pw] {\rxrzaccc};
            \node[anchor=south east, font=\bfseries] at (rel axis cs:0.95, 0.05) {RxRz};

        \nextgroupplot[xticklabels=\empty] 
            \addplot table[x=layer, y=lin] {\ryrzaccc};
            \addplot table[x=layer, y=full] {\ryrzaccc};
            \addplot table[x=layer, y=circ] {\ryrzaccc};
            \addplot table[x=layer, y=pw] {\ryrzaccc};
            \node[anchor=south east, font=\bfseries] at (rel axis cs:0.95, 0.05) {RyRz};
            
        \nextgroupplot[xticklabels=\empty] 
            \addplot table[x=layer, y=lin] {\rzryaccc};
            \addplot table[x=layer, y=full] {\rzryaccc};
            \addplot table[x=layer, y=circ] {\rzryaccc};
            \addplot table[x=layer, y=pw] {\rzryaccc};
            \node[anchor=south east, font=\bfseries] at (rel axis cs:0.95, 0.05) {RzRy};
            
        \end{groupplot}

            \node at (myplots c1r1.north west) [rotate=90, anchor=south, yshift=2em, xshift=-4em] {Accuracy $6$C $C_1$};

        \end{tikzpicture}
        \label{fig:class4_c1_sub}
    \end{subfigure}


    \begin{subfigure}{\textwidth}
        \centering
        \begin{tikzpicture}
            \begin{groupplot}[
                group style={
                    group size=4 by 2,
                    vertical sep=0.2cm,
                    horizontal sep=0.8cm,  
                    group name=myplots2,
                },
                width=0.17\columnwidth,    
                height=0.07\columnwidth,
                scale only axis,
                title={},
                xmin=1, xmax=5,
                xtick={1, 2, 3, 4, 5},
                samples at={1, 2, 3, 4, 5},
                ymin=0.3, ymax=1,
                ytick={0.3, 0.5, 0.7, 0.9},
                cycle list name=exotic,
                legend style={font=\tiny, at={(1.05,1)}, anchor=north west},
            ]

        
        \nextgroupplot[xticklabels=\empty] 
            \addplot table[x=layer, y=lin] {\rxaccxc};
            \addplot table[x=layer, y=full] {\rxaccxc};
            \addplot table[x=layer, y=circ] {\rxaccxc};
            \addplot table[x=layer, y=pw] {\rxaccxc};
            \node[anchor=north east, font=\bfseries] at (rel axis cs:0.95, 0.95) {Rx};

        \nextgroupplot[xticklabels=\empty] 
            \addplot table[x=layer, y=lin] {\rxryaccxc};
            \addplot table[x=layer, y=full] {\rxryaccxc};
            \addplot table[x=layer, y=circ] {\rxryaccxc};
            \addplot table[x=layer, y=pw] {\rxryaccxc};
            \node[anchor=south east, font=\bfseries] at (rel axis cs:0.95, 0.05) {RxRy};

        \nextgroupplot[xticklabels=\empty] 
            \addplot table[x=layer, y=lin] {\ryrxaccxc};
            \addplot table[x=layer, y=full] {\ryrxaccxc};
            \addplot table[x=layer, y=circ] {\ryrxaccxc};
            \addplot table[x=layer, y=pw] {\ryrxaccxc};
            \node[anchor=south east, font=\bfseries] at (rel axis cs:0.95, 0.05) {RyRx};
            
        \nextgroupplot[xticklabels=\empty] 
            \addplot table[x=layer, y=lin] {\rzrxaccxc}; \addlegendentry{Linear}
            \addplot table[x=layer, y=full] {\rzrxaccxc}; \addlegendentry{Full}
            \addplot table[x=layer, y=circ] {\rzrxaccxc}; \addlegendentry{Circular}
            \addplot table[x=layer, y=pw] {\rzrxaccxc}; \addlegendentry{Pairwise}
            \node[anchor=south east, font=\bfseries] at (rel axis cs:0.95, 0.05) {RzRx};
        

        \nextgroupplot 
            \addplot table[x=layer, y=lin] {\ryaccxc};
            \addplot table[x=layer, y=full] {\ryaccxc};
            \addplot table[x=layer, y=circ] {\ryaccxc};
            \addplot table[x=layer, y=pw] {\ryaccxc};
            \node[anchor=south east, font=\bfseries] at (rel axis cs:0.95, 0.05) {Ry};

        \nextgroupplot 
            \addplot table[x=layer, y=lin] {\rxrzaccxc};
            \addplot table[x=layer, y=full] {\rxrzaccxc};
            \addplot table[x=layer, y=circ] {\rxrzaccxc};
            \addplot table[x=layer, y=pw] {\rxrzaccxc};
            \node[anchor=south east, font=\bfseries] at (rel axis cs:0.95, 0.05) {RxRz};

        \nextgroupplot 
            \addplot table[x=layer, y=lin] {\ryrzaccxc};
            \addplot table[x=layer, y=full] {\ryrzaccxc};
            \addplot table[x=layer, y=circ] {\ryrzaccxc};
            \addplot table[x=layer, y=pw] {\ryrzaccxc};
            \node[anchor=south east, font=\bfseries] at (rel axis cs:0.95, 0.05) {RyRz};
            
        \nextgroupplot 
            \addplot table[x=layer, y=lin] {\rzryaccxc};
            \addplot table[x=layer, y=full] {\rzryaccxc};
            \addplot table[x=layer, y=circ] {\rzryaccxc};
            \addplot table[x=layer, y=pw] {\rzryaccxc};
            \node[anchor=south east, font=\bfseries] at (rel axis cs:0.95, 0.05) {RzRy};
            
        \end{groupplot}

            \node[anchor=north] at ($(myplots2 c2r2.south)!0.5!(myplots2 c3r2.south)$) [yshift=-1em, xshift=1em] {number of layers};
            \node at (myplots2 c1r1.north west) [rotate=90, anchor=south, yshift=2em, xshift=-4em] {Accuracy $6$C $C2$};

        \end{tikzpicture}
        \label{fig:class6_c1_sub}
    \end{subfigure}
    \captionsetup{skip=-0.05pt}
    \caption{Classification accuracy results of $C_1$ and $C_2$ with $6$ classes.}
    \label{fig:combined_class6}
\end{figure*}
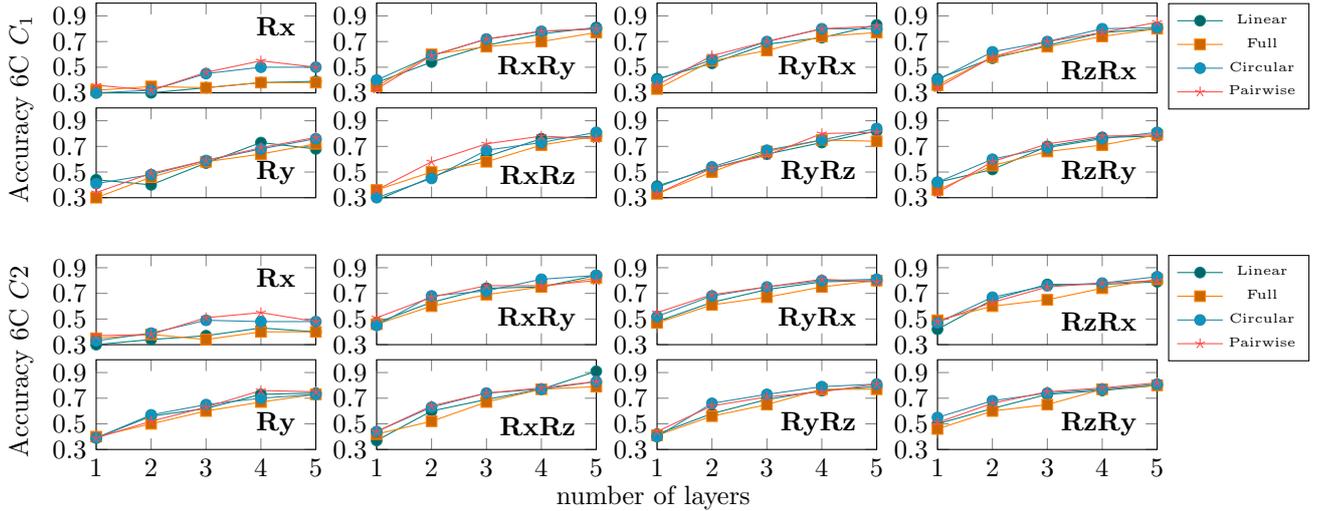

In Figures~\ref{fig:combined_fid},~\ref{fig:combined_hell},~\ref{fig:combined_class4} and~\ref{fig:combined_class6}, the results achieved by $C_1$ and $C_2$ on the various tasks are presented. The binary classification task is omitted, as all configurations have already been shown to easily achieve $99\%$ accuracy (see Table~\ref{tab:comparison2_3rots}). Adam is used (learning rate $0.01$, batch size $32$, and $5$ epochs) to train the network. To encode the images, amplitude embedding is used.

In the context of generation or classification tasks on the MNIST dataset, if  a low number of layers is considered, the best entanglement topologies are circular and pairwise, as they achieve the best performance, except for a single layer of $C_1$ and $C_2$ where the linear topology achieves comparable results in classification. 
However, as the number of layers increases, the performance saturates and all topologies perform similarly. This effect is particularly pronounced in the two-rotation combinations.

The generation of probability distributions works slightly better with the linear topology. However, all results are similar from the first layer, except for full topology, which starts to behave similarly from $2$ layers for $C_1$. All values are approximately $0.35$, with only minor variations.

Therefore, when the circuit is simple (a low number of layers) and the task is simple (generation of probability distributions), the linear entanglement performs very well. However, as the task becomes progressively more complex (such as the generation of images) or the circuit becomes more complex (a higher number of layers), the linear topology has deteriorated performance, while the circular and pairwise perform better. Moreover, this is confirmed by the classification task: when using the $C_1$ circuit with a single layer, the linear topology achieves the best performance along with the circular topology. However, as the number of layers increases, its performance deteriorates.

Therefore, across all tasks, the performance saturates in the same manner as the expressibility and entanglement values. Indeed, starting from $3$ layers, the expressibility and entanglement values begin to saturate, and similarly, the performance across the different tasks also saturates. This suggests that, as the number of layers increases, the choice of topology does not affect the performance; therefore, the topology with a lower number of gates should be used.
A similar effect seems to occur with the rotations. Starting from $3$
layers, the different rotation configurations reach a similar level of expressibility and entanglement and the same happens to the performance. Then, the choice of different rotations seems almost useless.

An additional interesting aspect emerges when the individual rotation configurations are considered with a low number of layers. In both image generation and classification, the single $R_y$ rotation performs better than the $R_x$ rotation, while in the probability distribution task, the $R_x$ rotation achieves better results. Therefore, this may suggest that with image-based tasks, $R_y$ performs better; however, this has to be investigated further, as it could be specific to the MNIST dataset. Furthermore, in image-based tasks, the pairwise topology seems to perform better. This could be due to its ability to connect neighboring qubits and, consequently, neighboring pixels.

Additionally, the combinations of $R_x$ and $R_y$ rotations generally perform better with a lower number of layers. This suggests that adding a phase using the $R_z$ rotation is not particularly useful or at least not as effective as the rotations on the x and y axes. The addition of a $R_x$ or $R_y$ gate always improves the performance, while the $R_z$ gate does not. It is understandable when $R_z$ is applied as the first rotation to the $\ket{0}$ state, as it only adds a phase to $\ket{0}$. 
However, there are cases where it performed well with image tasks. For instance, in the classification tasks, all combinations of the two rotations, except for the $R_xR_z$ rotational layer, performed similarly, although the $R_xR_y$ and $R_yR_x$ are the most consistent and they are the combinations with the highest and lowest entanglement values, respectively; thus, higher entanglement does not appear to be an issue, even though barren plateaus should be studied~\cite{ortiz2021entanglement}. 

Furthermore, when considering the circuit $C_1$ with a single rotation configuration and two layers, it leads to better expressibility values compared to the two-rotation configurations with only a single layer, which have the same number of parameters, although both configurations have similar entanglement values.
However, this higher expressibility of the two-layer circuits leads to worse performance with respect to the one-layer circuits in all tasks. Therefore, having a high expressibility is ineffective if the circuit does not also enhance its entanglement capability.

\pgfplotstableread{
layer   lin     full      circ      pw
1       200    300      150      180
2       200        230      130   130
3       200    170       130     130
4       160    130      95    110
5       130    130      100    120
}\rxfid

\pgfplotstableread{
layer   lin     full      circ      pw
1       175    250      150      160
2       145    170      120       115
3       95    90       90     85
4       70    80      50      70
5       65    70      60      65
}\ryfid

\pgfplotstableread{
layer   lin     full      circ      pw
1       150    260      120      140
2       100    110      55     50
3       55    60       50    40
4       40    50      40    40
5       35    40      40    40
}\rxryfid
\pgfplotstableread{
layer   lin     full      circ      pw
1       200     280      145      230
2       200    170      100     70
3       85  50       45    45
4       50   50      45    40
5       35   50      45    50

}\rxrzfid
\pgfplotstableread{
layer   lin     full      circ      pw
1       135     260        100      140
2       100    120      60     50
3       50    55       35     45
4       40   55      40    45
5       35   45      40    35
}\ryrxfid
\pgfplotstableread{
layer   lin     full      circ      pw
1       170    260      160      170
2       105    100      80     80
3       50   55       55    50
4       55  50      40    40
5       40  55      40    50
}\ryrzfid
\pgfplotstableread{
layer   lin     full      circ      pw
1       115    280      120      190
2       120    140      90     60
3       55   55       40    55
4       50   55      50    45
5       50   50      40    40
}\rzrxfid
\pgfplotstableread{
layer   lin     full      circ      pw
1       130    270      150      180
2       100    115       60     50
3       50    55       50     50
4       45    45      40    45
5       40    40      40    40
}\rzryfid

\pgfplotstableread{
layer   lin     full      circ      pw
1       180    160      130      160
2       180        160      95   95
3       180    145       95     100
4       130    130      95    80
5       110    130      100    100
}\rxfidc

\pgfplotstableread{
layer   lin     full      circ      pw
1       135    130      130      100
2       110    90      85       65
3       90    90       65     55
4       75    55      55      55
5       60    60      60      55
}\ryfidc

\pgfplotstableread{
layer   lin     full      circ      pw
1       95    100      70      70
2       60    90      50     60
3       45    65       55    40
4       40    50      45    40
5       35    40      40    40
}\rxryfidc
\pgfplotstableread{
layer   lin     full      circ      pw
1       115     130      90      90
2       90    80      60     45
3       70  45       45    50
4       50   45      45    40
5       35   50      45    50

}\rxrzfidc
\pgfplotstableread{
layer   lin     full      circ      pw
1       120     110        100      70
2       55    50      55     40
3       40    45       40     40
4       40   40      40    45
5       35   45      40    35
}\ryrxfidc
\pgfplotstableread{
layer   lin     full      circ      pw
1       100    120      75      85
2       90    65      45     40
3       40   45       40    40
4       50  50      40    40
5       40  55      40    50
}\ryrzfidc
\pgfplotstableread{
layer   lin     full      circ      pw
1       85    110      55      70
2       60    70      40     40
3       45   45       45    45
4       50   50      45    45
5       50   50      40    40
}\rzrxfidc
\pgfplotstableread{
layer   lin     full      circ      pw
1       120    110      80      70
2       90    50       45     50
3       45    45       45     40
4       40    40      40    45
5       40    40      40    40
}\rzryfidc

\pgfplotstableread{
layer   lin     full      circ      pw
1       0.31    0.42      0.39      0.44
2       0.42        0.4      0.35   0.37
3       0.42    0.42       0.34     0.34
4       0.39    0.39      0.28    0.32
5       0.39    0.3      0.29    0.3
}\rxhell

\pgfplotstableread{
layer   lin     full      circ      pw
1       0.35    0.42      0.35      0.43
2       0.41    0.55      0.35       0.4
3       0.41    0.5       0.4     0.43
4       0.35    0.5      0.4      0.4
5       0.35    0.45      0.4      0.42
}\ryhell

\pgfplotstableread{
layer   lin     full      circ      pw
1       0.32    0.38      0.33      0.4
2       0.27    0.37      0.32     0.33
3       0.3    0.4       0.32    0.33
4       0.35    0.38      0.35    0.28
5       0.28    0.34      0.35    0.31
}\rxryhell
\pgfplotstableread{
layer   lin     full      circ      pw
1       0.31     0.41      0.35      0.45
2       0.39    0.41      0.39     0.42
3       0.4  0.38       0.39    0.38
4       0.27   0.34      0.39    0.35
5       0.32   0.34      0.36    0.29

}\rxrzhell
\pgfplotstableread{
layer   lin     full      circ      pw
1       0.32     0.36        0.35      0.4
2       0.28    0.41      0.31     0.46
3       0.3    0.39       0.29     0.32
4       0.33   0.42      0.35    0.29
5       0.27   0.3      0.33    0.28
}\ryrxhell
\pgfplotstableread{
layer   lin     full      circ      pw
1       0.31    0.42      0.38      0.45
2       0.41    0.49      0.38     0.42
3       0.36   0.47       0.4    0.36
4       0.4  0.46      0.33    0.3
5       0.3  0.37      0.35    0.31
}\ryrzhell
\pgfplotstableread{
layer   lin     full      circ      pw
1       0.33    0.42      0.4      0.44
2       0.46    0.49      0.4     0.4
3       0.4   0.4       0.38    0.34
4       0.35   0.3      0.35    0.4
5       0.27   0.3      0.31    0.31
}\rzrxhell
\pgfplotstableread{
layer   lin     full      circ      pw
1       0.33    0.41      0.33      0.45
2       0.3    0.43       0.33     0.39
3       0.33    0.45       0.38     0.38
4       0.33    0.45      0.33    0.32
5       0.34    0.4      0.34    0.3
}\rzryhell

\pgfplotstableread{
layer   lin     full      circ      pw
1       0.4    0.39      0.41      0.42
2       0.44        0.38      0.36   0.4
3       0.37    0.34       0.33     0.35
4       0.3    0.32      0.32    0.38
5       0.33    0.31      0.32    0.37
}\rxhellc

\pgfplotstableread{
layer   lin     full      circ      pw
1       0.59    0.53      0.53      0.61
2       0.53    0.44      0.45       0.47
3       0.41    0.5       0.39     0.48
4       0.37    0.48      0.4      0.41
5       0.4    0.4      0.4      0.38
}\ryhellc

\pgfplotstableread{
layer   lin     full      circ      pw
1       0.3    0.37      0.3      0.35
2       0.34    0.39      0.3     0.31
3       0.30    0.36       0.31    0.33
4       0.3    0.33      0.3    0.32
5       0.31    0.34      0.28    0.32
}\rxryhellc
\pgfplotstableread{
layer   lin     full      circ      pw
1       0.24     0.34      0.24      0.34
2       0.34    0.33      0.31     0.38
3       0.32  0.29       0.32    0.35
4       0.31       0.27   0.37      0.3   
5       0.34 0.3  0.28   0.3
}\rxrzhellc
\pgfplotstableread{
layer   lin     full      circ      pw
1       0.29     0.37        0.32      0.33
2       0.33    0.44      0.28     0.3
3       0.32    0.32       0.29     0.35
4       0.3   0.33      0.31    0.33
5       0.32   0.32      0.31    0.31
}\ryrxhellc
\pgfplotstableread{
layer   lin     full      circ      pw
1       0.49    0.44      0.46      0.53
2       0.43    0.44      0.36     0.42
3       0.34   0.37       0.31    0.36
4       0.31  0.36      0.33    0.36
5       0.29  0.31      0.3    0.3
}\ryrzhellc
\pgfplotstableread{
layer   lin     full      circ      pw
1       0.35    0.36      0.35      0.33
2       0.42    0.35      0.35     0.32
3       0.31   0.33       0.29    0.36
4       0.33   0.38      0.32    0.34
5       0.34   0.32      0.3    0.4
}\rzrxhellc
\pgfplotstableread{
layer   lin     full      circ      pw
1       0.47    0.44      0.42      0.51
2       0.42    0.38       0.33     0.38
3       0.38    0.35       0.38     0.36
4       0.32    0.4      0.33    0.32
5       0.4    0.36      0.31    0.34
}\rzryhellc

\pgfplotstableread{
layer   lin     full      circ      pw
1       0.5    0.5      0.5      0.5
2       0.5        0.5      0.55   0.53
3       0.5    0.5       0.65     0.63
4       0.68    0.67      0.71    0.6
5       0.6    0.6      0.68    0.62
}\rxacc

\pgfplotstableread{
layer   lin     full      circ      pw
1       0.6    0.5      0.6      0.5
2       0.68    0.67      0.7       0.66
3       0.8    0.74       0.76     0.83
4       0.85    0.79      0.83      0.85
5       0.89    0.88      0.89      0.88
}\ryacc

\pgfplotstableread{
layer   lin     full      circ      pw
1       0.6    0.5      0.6      0.5
2       0.75    0.7      0.77     0.76
3       0.85    0.80       0.86    0.87
4       0.91    0.85      0.92    0.89
5       0.90    0.91      0.90    0.88
}\rxryacc
\pgfplotstableread{
layer   lin     full      circ      pw
1       0.5     0.5      0.5      0.5
2       0.65    0.7      0.78     0.73
3       0.72  0.75       0.84    0.84
4       0.85   0.85      0.87    0.90
5       0.89   0.90      0.91    0.90

}\rxrzacc
\pgfplotstableread{
layer   lin     full      circ      pw
1       0.6     0.5        0.6      0.5
2       0.73    0.73      0.77     0.77
3       0.83    0.76       0.84     0.85
4       0.89   0.87      0.90    0.86
5       0.89   0.88      0.92    0.91
}\ryrxacc
\pgfplotstableread{
layer   lin     full      circ      pw
1       0.6    0.5      0.6      0.5
2       0.65    0.68      0.74     0.72
3       0.83   0.74       0.85    0.82
4       0.85  0.80      0.87    0.89
5       0.90  0.90      0.90    0.91
}\ryrzacc
\pgfplotstableread{
layer   lin     full      circ      pw
1       0.6    0.5      0.6      0.5
2       0.73    0.68      0.77     0.73
3       0.85   0.80       0.85    0.85
4       0.88   0.89      0.88    0.89
5       0.91   0.88      0.90    0.90
}\rzrxacc
\pgfplotstableread{
layer   lin     full      circ      pw
1       0.6    0.5      0.6      0.5
2       0.74    0.74       0.77     0.80
3       0.85    0.80       0.87     0.85
4       0.88    0.88      0.85    0.90
5       0.91    0.91      0.92    0.90
}\rzryacc

\pgfplotstableread{
layer   lin     full      circ      pw
1       0.5    0.5      0.52      0.5
2       0.5        0.5      0.62   0.66
3       0.5    0.5       0.62     0.65
4       0.67    0.68      0.7    0.65
5       0.6    0.6      0.6    0.62
}\rxaccx

\pgfplotstableread{
layer   lin     full      circ      pw
1       0.66    0.5      0.61      0.58
2       0.7    0.7      0.74       0.78
3       0.8    0.76       0.84     0.82
4       0.84    0.83      0.9      0.86
5       0.9    0.85      0.90      0.86
}\ryaccx

\pgfplotstableread{
layer   lin     full      circ      pw
1       0.64    0.6      0.73      0.75
2       0.81    0.76      0.84     0.82
3       0.87    0.81       0.88    0.88
4       0.9 0.89      0.92    0.90
5       0.91    0.89      0.92    0.92
}\rxryaccx
\pgfplotstableread{
layer   lin     full      circ      pw
1       0.52     0.5      0.64      0.58
2       0.77    0.72      0.80     0.78
3       0.83  0.77       0.87    0.87
4       0.9       0.88   0.9      0.92   
5       0.92 0.89  0.93   0.92
}\rxrzaccx
\pgfplotstableread{
layer   lin     full      circ      pw
1       0.64     0.56        0.66      0.61
2       0.77    0.75      0.83     0.83
3       0.86    0.8       0.87     0.88
4       0.91   0.86      0.88    0.91
5       0.92   0.90      0.93    0.91
}\ryrxaccx
\pgfplotstableread{
layer   lin     full      circ      pw
1       0.59    0.55      0.67      0.64
2       0.76    0.72      0.79     0.79
3       0.85   0.77       0.87    0.86
4       0.90  0.86      0.91    0.89
5       0.91  0.86      0.92    0.91
}\ryrzaccx
\pgfplotstableread{
layer   lin     full      circ      pw
1       0.60    0.56      0.65      0.65
2       0.82    0.75      0.83     0.82
3       0.87   0.81       0.87    0.85
4       0.90   0.85      0.92    0.89
5       0.92   0.89      0.91    0.93
}\rzrxaccx
\pgfplotstableread{
layer   lin     full      circ      pw
1       0.6    0.56      0.65      0.67
2       0.80    0.78       0.82     0.82
3       0.87    0.82       0.86     0.88
4       0.89    0.87      0.90    0.90
5       0.91    0.89      0.91    0.92
}\rzryaccx

\pgfplotstableread{
layer   lin     full      circ      pw
1       0.3    0.32      0.3      0.36
2       0.3        0.35      0.32   0.32
3       0.34    0.34       0.45     0.46
4       0.38    0.38      0.5    0.55
5       0.39    0.38      0.5    0.5
}\rxaccc

\pgfplotstableread{
layer   lin     full      circ      pw
1       0.44    0.3      0.41      0.34
2       0.4    0.46      0.48       0.49
3       0.57    0.58       0.59     0.59
4       0.73    0.64      0.68      0.69
5       0.68    0.72      0.76      0.77
}\ryaccc

\pgfplotstableread{
layer   lin     full      circ      pw
1       0.38    0.35      0.4      0.33
2       0.54    0.6      0.59     0.59
3       0.67    0.66       0.72    0.72
4       0.76    0.7      0.78    0.78
5       0.81    0.77      0.8    0.8
}\rxryaccc
\pgfplotstableread{
layer   lin     full      circ      pw
1       0.28     0.36      0.3      0.36
2       0.46    0.5      0.45     0.58
3       0.62  0.58       0.67    0.72
4       0.76   0.71      0.73    0.78
5       0.78   0.78      0.81    0.76

}\rxrzaccc
\pgfplotstableread{
layer   lin     full      circ      pw
1       0.41     0.33        0.4      0.36
2       0.53    0.55      0.56     0.59
3       0.68    0.63       0.7     0.7
4       0.73   0.74      0.8    0.8
5       0.83   0.77      0.8    0.82
}\ryrxaccc
\pgfplotstableread{
layer   lin     full      circ      pw
1       0.39    0.33      0.38      0.33
2       0.53    0.5      0.54     0.53
3       0.64   0.66       0.67    0.63
4       0.73  0.75      0.75    0.8
5       0.82  0.74      0.84    0.81
}\ryrzaccc
\pgfplotstableread{
layer   lin     full      circ      pw
1       0.41    0.36      0.4      0.34
2       0.57    0.58      0.62     0.58
3       0.67 0.66       0.7    0.7
4       0.77   0.74      0.8    0.77
5       0.80   0.8      0.81    0.85
}\rzrxaccc
\pgfplotstableread{
layer   lin     full      circ      pw
1       0.42    0.36      0.42      0.34
2       0.52    0.55       0.60     0.58
3       0.7    0.66       0.69     0.72
4       0.77    0.71      0.76    0.78
5       0.78    0.79      0.81    0.79
}\rzryaccc

\pgfplotstableread{
layer   lin     full      circ      pw
1       0.3    0.35      0.33      0.37
2       0.34        0.38      0.39   0.38
3       0.37    0.34       0.49     0.51
4       0.43    0.4      0.48    0.55
5       0.4    0.4      0.48    0.48
}\rxaccxc

\pgfplotstableread{
layer   lin     full      circ      pw
1       0.39    0.4      0.39      0.39
2       0.56    0.5      0.57       0.52
3       0.62    0.6       0.65     0.63
4       0.73    0.67      0.7      0.76
5       0.74    0.73      0.73      0.75
}\ryaccxc

\pgfplotstableread{
layer   lin     full      circ      pw
1       0.47    0.46      0.45      0.51
2       0.63    0.6      0.68     0.67
3       0.74    0.69       0.72    0.76
4       0.75    0.75      0.81    0.76
5       0.84    0.82      0.84    0.8
}\rxryaccxc
\pgfplotstableread{
layer   lin     full      circ      pw
1       0.37     0.42      0.44      0.44
2       0.6    0.52      0.63     0.64
3       0.69  0.67       0.74    0.74
4       0.77       0.77   0.77      0.78   
5       0.91 0.79  0.83   0.83
}\rxrzaccxc
\pgfplotstableread{
layer   lin     full      circ      pw
1       0.48     0.47        0.52      0.55
2       0.63    0.61      0.68     0.69
3       0.73    0.67       0.75     0.75
4       0.79   0.75      0.8    0.81
5       0.80   0.80      0.81    0.79
}\ryrxaccxc
\pgfplotstableread{
layer   lin     full      circ      pw
1       0.42    0.41      0.4      0.44
2       0.58    0.56      0.66     0.64
3       0.69   0.65       0.73    0.71
4       0.76  0.77      0.79    0.75
5       0.79  0.77      0.81    0.81
}\ryrzaccxc
\pgfplotstableread{
layer   lin     full      circ      pw
1       0.42    0.49      0.47      0.48
2       0.65    0.6      0.67     0.63
3       0.77   0.65       0.76    0.75
4       0.77   0.74      0.78    0.78
5       0.79   0.81      0.83    0.8
}\rzrxaccxc
\pgfplotstableread{
layer   lin     full      circ      pw
1       0.5    0.46      0.55      0.51
2       0.62    0.6       0.68     0.66
3       0.73    0.65       0.74     0.75
4       0.76    0.77      0.77    0.78
5       0.8    0.8      0.81    0.82
}\rzryaccxc

\section{Execution on Real Quantum Computers}
\label{results_ibm}
The best-performing circuits between the execution with a low number of layers for each task are also tested on real IBM hardware to assess their performance. In particular, the generation of probability distributions is tested with $R_x$ rotation and with linear topology with only a single layer of $C_1$. The generation of images is tested with $R_xR_y$ combinations, with the circular topology and two layers. The 2/4-class classifications are tested, respectively, with: a single layer of the $C_1$ circuit with an $R_y$ gate and circular topology, and two layers of the $C_1$ circuit with $R_xR_y$ gates and circular topology.

Tables~\ref{tab:resultsIBM} and~\ref{tab:resultsIBM2} report the QNN results. The tests are performed with 1024 shots. The MNIST testset is used to assess classification performance, while 30 random probability distributions and 100 fake images are used for QGAN tasks. The performance achieved on the real hardware is quite similar to that obtained in simulation, except for the classification task. In particular, the probability distribution task achieves a Hellinger distance of $0.35$ (vs. $0.31$ in simulation). The generation of images obtains a FID score of $80$ (vs. $55$). In contrast, classification accuracy drops significantly: $56\%$ (vs. $99\%$) for two classes, and $40\%$ (vs. $78\%$) for four. This discrepancy is mainly due to the amplitude encoding used to map images into quantum states. This encoding method significantly increases the depth of the quantum circuit compared to the two generative tasks.

\begin{table}[htbp]
\centering
\resizebox{0.45\textwidth}{!}{
\renewcommand{\arraystretch}{1.2}
    \begin{tabular}{|>{\centering\arraybackslash}m{1cm}||>{\centering\arraybackslash}m{0.8cm}|>{\centering\arraybackslash}m{1cm}|>{\centering\arraybackslash}m{1cm}|>{\centering\arraybackslash}m{0.8cm}|>{\centering\arraybackslash}m{0.8cm}|>{\centering\arraybackslash}m{0.8cm}|>{\centering\arraybackslash}m{0.8cm}|}
     \hline
     Task & Circuit &Rot Gate &$\#$ layer & Ent topology & IBM Fez & IBM Marrakesh & IBM Kingston  \\
     \hline
     QGAN Probs&$C_1$&$R_x$ & 1 & Linear &$0.34$ &$0.36$&$0.35$ \\
     \hline
     2 Classes&$C_1$&$R_y$ & 1 & Circular&$56\%$&$50\%$&$47\%$\\
     \hline
    \end{tabular}
    }
    \caption{Results on real IBM quantum computers. Metrics: Hellinger distance for QGAN and accuracy for classification.}
    \label{tab:resultsIBM}
\end{table}

\begin{table}[htbp]
\centering
\resizebox{0.38\textwidth}{!}{
\renewcommand{\arraystretch}{1.2}
    \begin{tabular}{|>{\centering\arraybackslash}m{1cm}||>{\centering\arraybackslash}m{0.8cm}|>{\centering\arraybackslash}m{1cm}|>{\centering\arraybackslash}m{1cm}|>{\centering\arraybackslash}m{0.8cm}|>{\centering\arraybackslash}m{0.8cm}|}
     \hline
     Task & Circuit &Rot Gate &$\#$ layer & Ent topology & IBM Marrakesh  \\
     \hline
     QGAN Images&$C_1$&$R_xR_y$ & 2 &Circular &$80$ \\
     \hline
     4 Classes&$C_1$&$R_xR_y$ & 2 & Circular&$40\%$\\
     \hline
    \end{tabular}
    }
    \caption{Results on real IBM quantum computer. Metrics: FID for QGAN and accuracy for classification.}
    \label{tab:resultsIBM2}
\end{table}

\section{Conclusion}
\label{conclusion}
In this work, an analysis of the performance of various VQCs is performed using two types of circuits: one with alternating rotations and entanglement layers, and another with an additional final rotations layer. The investigation explores different entanglement topologies (linear, circular, pairwise, and full), gate configurations (single and two-rotations), QML tasks (probability distribution generation, image generation, classification), and the number of layers.

The performance achieved in the three tasks is correlated with entanglement and expressibility. With few layers, $R_xR_y$ and $R_yR_x$ configurations consistently perform best, and with circular or pairwise topologies, while the full topology performs worse.  
Furthermore, when evaluating the circuits as a whole, with a low number of layers, the performance appears to depend more on the entanglement topology. Linear and full topologies exhibit lower entanglement and expressibility values and perform poorly, while circular and pairwise have a slightly higher entanglement and expressibility and achieve better results. However, as the number of layers increases, all topologies achieve comparable expressibility and entanglement, leading to similar performance saturation across all tasks. The same trend is observed with the rotations: as the number of layers increases, the performance of the different two-rotation configurations is similar and saturates, such as expressibility and entanglement.
Moreover, when using two-rotation configurations, with a low number of layers, considering the entanglement topologies separately, the best results are obtained with either the highest or lowest entanglement or expressibility.
Additionally, each task is tested on real IBM hardware to assess performance.

Future work will validate these findings across diverse tasks and datasets to assess their generalizability. The impact of Barren Plateaus will be considered. Furthermore, evaluating circuits with more qubits will help to verify the consistency of the results. Moreover, exploring mixed-gate layers and different entangling gates may also clarify the dependence of performance on specific gate choices.

\section*{Acknowledgment}
We acknowledge the financial support from Spoke 10 - ICSC - ``National Research Centre in High Performance Computing, Big Data and Quantum Computing'', funded by European Union – NextGenerationEU.
This research benefits from the High Performance Computing facility of the University of Parma, Italy (HPC.unipr.it) and also from IBM Quantum Credits awarded to Michele Amoretti - project Crosstalk-aware Quantum Multiprogramming.





\FloatBarrier
\bibliographystyle{unsrt}
\bibliography{references}

\begin{thebibliography}{10}

\bibitem{sim2019expressibility}
Sukin Sim, Peter~D Johnson, and Al{\'a}n Aspuru-Guzik.
\newblock Expressibility and entangling capability of parameterized quantum
  circuits for hybrid quantum-classical algorithms.
\newblock {\em Advanced Quantum Technologies}, 2(12):1900070, 2019.

\bibitem{holmes2022connecting}
Zo{\"e} Holmes, Kunal Sharma, Marco Cerezo, and Patrick~J Coles.
\newblock Connecting ansatz expressibility to gradient magnitudes and barren
  plateaus.
\newblock {\em PRX quantum}, 3(1):010313, 2022.

\bibitem{ortiz2021entanglement}
Carlos Ortiz~Marrero, M{\'a}ria Kieferov{\'a}, and Nathan Wiebe.
\newblock Entanglement-induced barren plateaus.
\newblock {\em PRX quantum}, 2(4):040316, 2021.

\bibitem{kim2022entanglement}
Joonho Kim and Yaron Oz.
\newblock Entanglement diagnostics for efficient vqa optimization.
\newblock {\em Journal of Statistical Mechanics: Theory and Experiment},
  2022(7):073101, 2022.

\bibitem{hubregtsen2021evaluation}
Thomas Hubregtsen, Josef Pichlmeier, Patrick Stecher, and Koen Bertels.
\newblock Evaluation of parameterized quantum circuits: on the relation between
  classification accuracy, expressibility, and entangling capability.
\newblock {\em Quantum Machine Intelligence}, 3(1):9, 2021.

\bibitem{ballarin2023entanglement}
Marco Ballarin, Stefano Mangini, Simone Montangero, Chiara Macchiavello, and
  Riccardo Mengoni.
\newblock Entanglement entropy production in quantum neural networks.
\newblock {\em Quantum}, 7:1023, 2023.

\bibitem{correr2024characterizing}
Guilherme~Il{\'a}rio Correr, Ivan Medina, Pedro~C Azado, Alexandre Drinko, and
  Diogo~O Soares-Pinto.
\newblock Characterizing randomness in parameterized quantum circuits through
  expressibility and average entanglement.
\newblock {\em Quantum Science and Technology}, 10(1):015008, 2024.

\bibitem{zoufal2019quantum}
Christa Zoufal, Aur{\'e}lien Lucchi, and Stefan Woerner.
\newblock Quantum generative adversarial networks for learning and loading
  random distributions.
\newblock {\em npj Quantum Information}, 5(1):103, 2019.

\bibitem{huang2021experimental}
He-Liang Huang, Yuxuan Du, Ming Gong, Youwei Zhao, Yulin Wu, Chaoyue Wang,
  Shaowei Li, Futian Liang, Jin Lin, Yu~Xu, et~al.
\newblock Experimental quantum generative adversarial networks for image
  generation.
\newblock {\em Physical Review Applied}, 16(2):024051, 2021.

\bibitem{meyer2001global}
David~A Meyer and Nolan~R Wallach.
\newblock Global entanglement in multiparticle systems.
\newblock {\em arXiv preprint quant-ph/0108104}, 2001.

\bibitem{brennen2003observable}
Gavin~K Brennen.
\newblock An observable measure of entanglement for pure states of multi-qubit
  systems.
\newblock {\em arXiv preprint quant-ph/0305094}, 2003.

\end{thebibliography}

\end{document}